\documentclass{vldb}
\usepackage{amssymb}
\usepackage{times}
\usepackage{latexsym, amssymb,epstopdf}
\usepackage{amsmath}
\usepackage{color}
\usepackage{url}
\usepackage{verbatim} 
\usepackage{moreverb}
\usepackage{graphicx}
\usepackage[caption=false,font=footnotesize]{subfig}
\usepackage{algorithm}
\usepackage{algorithmicx}
\usepackage[noend]{algpseudocode}
\usepackage{mdwlist}

\newenvironment{descriptionsmallermarginnobfnoem} {
\begin{basedescript}{\desclabelwidth{.15in}\setlength{\topsep}{2pt}\setlength{\itemsep}{0pt}  } }
{\end{basedescript}}

\newcommand{\topic}[1]{\vspace{4pt} \noindent \underline{\bf #1}}

\newcommand{\amolnote}[1]{\red{Amol: #1}}

\long\def\ignore#1{}

\newcommand{\red}[1]{\textcolor{red}{#1}}

\title{Efficient Snapshot Retrieval over Historical Graph Data}

\numberofauthors{2} 

\author{
\alignauthor
Udayan Khurana \\
       \affaddr{University of Maryland, College Park}\\
       \email{udayan@cs.umd.edu}
\alignauthor
Amol Deshpande \\
       \affaddr{University of Maryland, College Park}\\
       \email{amol@cs.umd.edu}
}
\newcounter{foo}
\newenvironment{myenumerate}{\begin{list}{\arabic{foo}.}{
        \usecounter{foo}
        \setlength{\leftmargin}{15pt}
        \setlength{\topsep}{2pt}
        \setlength{\itemsep}{1pt}
}}{ \end{list}}

\begin{document}
\maketitle

\begin{abstract}
We address the problem of managing historical data for large evolving information networks like
social networks or citation networks, with the goal to enable temporal and evolutionary queries and
analysis.
We present the design and architecture of a \textit{distributed} graph database system that stores the entire history of a network and provides support for
efficient retrieval of multiple graphs from arbitrary time points in the past, in addition to
maintaining the current state for ongoing updates. 
Our system exposes a general programmatic API to process and analyze the retrieved snapshots. 
We introduce \textit{DeltaGraph}, a
novel, extensible, highly tunable, and distributed hierarchical index structure that enables compactly recording the
historical information, and that supports efficient retrieval of historical graph snapshots for single-site
or parallel processing. 
Along with the original graph data, DeltaGraph can also maintain and index {\em auxiliary} 
information; this functionality can be used to extend the structure to efficiently execute queries like {\em subgraph pattern 
matching} over historical data.
We develop analytical models for both the storage space needed and the snapshot retrieval times to aid
in choosing the right parameters for a specific scenario. In addition, we present strategies for materializing
portions of the historical graph state in memory to further speed up the retrieval process. 
Secondly, we present an in-memory graph data structure called \textit{GraphPool} that can maintain hundreds of 
historical graph instances in main memory in a non-redundant manner. 
We present a comprehensive experimental evaluation 
that illustrates
the effectiveness of our proposed techniques at managing
historical graph information.

\end{abstract}

\section{Introduction}
In recent years, we have witnessed an increasing abundance of observational data describing various
types of information networks, including social networks, biological networks, citation networks,
financial transaction networks, communication networks, to name a few. There is much work on
analyzing such networks to understand various social and natural phenomena like: 
\textit{``how the entities in a network interact''}, \textit{``how information spreads''},
\textit{``what are the most
important ({\em central}) entities''}, and {\em ``what are the key building blocks of a network''}. 
With the increasing availability of the digital trace of such networks {\em over time}, the topic of
network analysis has naturally extended its scope to {\em temporal} analysis of networks, which has the
potential to lend much better insights into various phenomena, especially those relating to the
temporal or evolutionary aspects of the network. For example, we may want to know:
\textit{``which analytical model best captures network evolution''}, \textit{``how information spreads over time''}, 
\textit{``who are the people with the steepest
increase in centrality measures over a period of time''}, \textit{``what is the average monthly
density of the network since 1997''}, \textit{``how the clusters in the network
evolve over time''} etc. 
Historical queries like, \textit{``who had the highest
PageRank centrality in a citation network in 1960''}, \textit{``which year amongst 2001 and 2004
had the smallest network diameter''}, \textit{``how many new triangles have been formed in the
network over the last year''}, also involve the temporal aspect of the network. More generally a
network analyst may want to process the historical trace of a network in different, usually
unpredictable, ways to gain insights into various phenomena. There is also interest in
visualizations of such networks over time~\cite{AhnTaxonomy2011}.

To support a broad range of network analysis tasks, we require a graph\footnote{We use the
terms graph and network interchangeably in this paper.} data management
system at the backend capable of low-cost storage and efficient retrieval of the historical network
information, in addition to maintaining the \emph{current} state of the network for updates and
other queries, temporal or otherwise. However, the existing solutions for graph data management lack
adequate techniques for temporal annotation, or for storage and retrieval of large scale historical
changes on the graph. In this paper, we present the design of a graph data management system that we
are building to provide support for executing temporal analysis queries and historical queries over
large-scale evolving information networks. 

Our focus in this paper is on supporting {\bf snapshot
queries} where the goal is to retrieve in memory one or more historical snapshots of the information network as
of specified time points. The typically unpredictable and non-declarative nature of network
analysis makes this perhaps the most important type of query that needs to be supported by such a
system.  Furthermore, the snapshot retrieval times must be sufficiently
low so that the analysis can be performed in an interactive manner (this is especially vital for
visualization tools). There is a large body of work in temporal relational databases that has
attempted to address some of these challenges for relational data (we discuss the related work in
more detail in the next section). However our primary focus on efficiently retrieving possibly 100's of 
snapshots in memory, while maintaining the current state of the database for ongoing updates and
queries, required us to rethink several key design decisions and data structures in such a system.  We assume there is enough memory to hold the retrieved snapshots in memory (we discuss below how 
we exploit overlap in the retrieved snapshots to minimize the memory requirements); we allow the snapshots
to be retrieved in a {\bf partitioned} fashion across a set of machines in parallel to handle very large scale networks. This
design decision was motivated by both the current hardware trends and the fact that, most network analysis tasks tend to 
access the underlying network in unpredictable ways, leading to unacceptably high penalties if the data does not 
fit in memory. Most current large-scale graph analysis systems, including Pregel~\cite{pregel}, Giraph\footnote{\url{http://giraph.apache.org}}, Trinity~\cite{trinity},
Cassovary (Twitter graph library)\footnote{\url{https://github.com/twitter/cassovary}}, Pegasus~\cite{Kang:2009:PPG:1674659.1677058}, load the entire graph into memory 
prior to execution.

\begin{figure}[t]
\centering
\includegraphics [width=0.23\textwidth,height=2in]{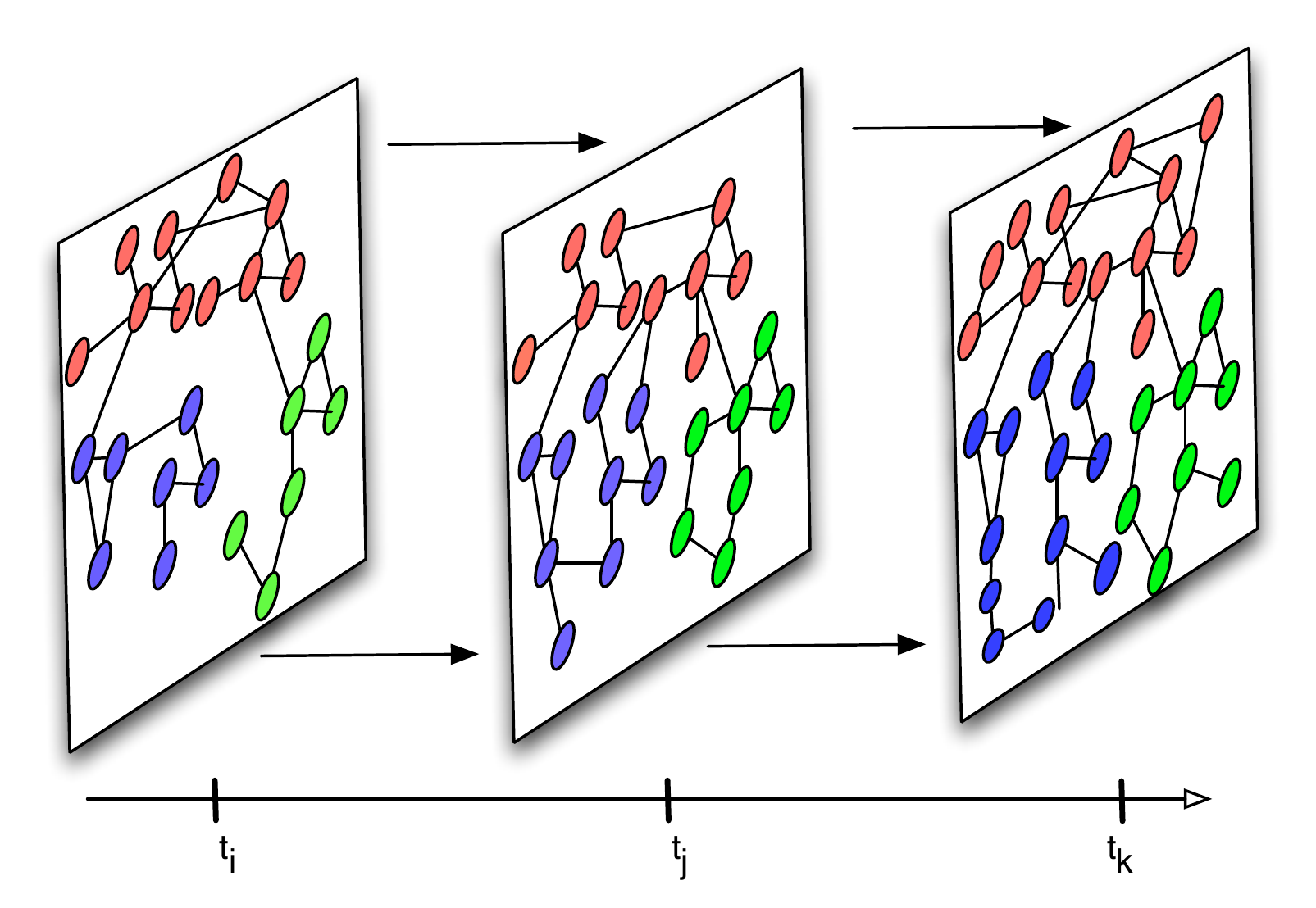}
\includegraphics [width=0.23\textwidth]{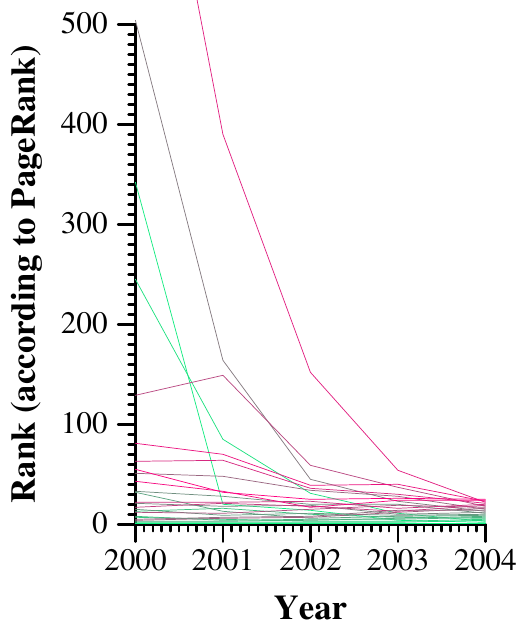}
\caption{Dynamic network analysis (e.g., understanding how ``communities'' evolve in a social network, how centrality scores
change in co-authorship networks, etc.) can 
lend important insights into social, cultural, and natural phenomena. The right plot was constructed
using our system over the DBLP network, and shows the evolution of the nodes ranked in top 25 in 2004.}
\label{fig:intro1}
\end{figure}

The cornerstone of our system is a novel hierarchical index structure, called {\em DeltaGraph}, over the
historical trace of a network. A DeltaGraph is a rooted, directed graph whose lowest level corresponds to the 
snapshots of the network over time (that are not explicitly stored), and the interior nodes
correspond to graphs constructed by combining the lower level graphs (these are typically not valid
graphs as of any specific time point). The information stored with each edge, called {\em edge deltas}, 
is sufficient to construct the graph corresponding to the target node from the graph
corresponding to the source node, and thus a specific snapshot can be created by traversing any path
from the root to the snapshot. While conceptually simple, DeltaGraph is a very powerful, extensible, 
general, and tunable index structure that enables trading off the different resources and
user requirements as per a specific application's need. By appropriately tuning the DeltaGraph construction
parameters, we can trade decreased snapshot retrieval times for increased disk storage requirements.
One key parameter of the DeltaGraph enables us to control the distribution of average
snapshot retrieval times over history. 
Portions of the DeltaGraph can be pre-fetched and {\em materialized}, allowing us to trade increased
memory utilization for reduced query times.  Many of these decisions can be made at run-time,
enabling us to adaptively respond to changing resource characteristics or user requirements. One
immediate consequence is also that we can optimally use the ``current graph'' that is in memory at all times
to reduce the snapshot retrieval times. DeltaGraph is highly extensible, providing a user the opportunity to define additional 
indexes to be stored on edge deltas in order to perform specific operations more efficiently.  Finally, DeltaGraph utilizes several other optimizations, including a 
{\em column-oriented storage} to minimize the amount of data that needs to be fetched to answer
a query, and multi-query optimization to simultaneously retrieve many snapshots.

DeltaGraph naturally enables distributed storage and processing to scale to very large graphs. The edge
deltas can be stored in a distributed fashion through use of horizontal partitioning, and the historical 
snapshots can be loaded parallely onto a set of machines in a partitioned fashion; in general, the two 
partitionings need not be aligned, but for computational efficiency, we currently require that they be 
aligned. Horizontal partitioning also results in lower snapshot retrieval latencies since the 
different deltas needed for reconstruction can be fetched in parallel.

The second key component of our system is an in-memory data structure called {\em GraphPool}.
A typical network evolution query may require analyzing 100's of snapshots from the history of a graph. 
Maintaining these snapshots in memory independently of each other would likely be infeasible. 
The GraphPool data structure exploits the commonalities in the snapshots that are currently in
memory, by overlaying them on a single graph data structure (typically a {\em union} of all the snapshots in memory).
GraphPool also employs several optimizations to minimize the amount of work needed to incorporate a
new snapshot and to clean up when a snapshot is purged after the analysis has completed.
We have implemented the GraphPool as a completely in-memory structure because of the inefficiency of general graph 
algorithms on a disk-resident/buffered-memory storage scheme. However, the design of our system itself 
enforces no such restriction on GraphPool. We address the general problem of scalability through horizontal 
partitioning, as explained later.

We have built a prototype implementation of our system in Java, using the Kyoto 
Cabinet\footnote{\url{http://fallabs.com/kyotocabinet}}
disk-based key-value store as the back-end engine to store the DeltaGraph; in the distributed case, we
run one instance on each machine. Our design decision to use a
key-value store at the back-end was motivated by the flexibility, the fast retrieval times, and the scalability 
afforded by such systems; since we only require a simple {\em get/put} interface from the storage engine, 
we can easily plug in other cloud-based, distributed key-value stores like 
HBase\footnote{\url{http://hbase.apache.org}} or Cassandra~\cite{cassandra}. Our comprehensive
experimental evaluation shows that our system can retrieve historical snapshots containing up to
millions of nodes and edges in several 100's of milliseconds or less, often an order of magnitude faster
than prior techniques like {\em interval trees}, and that the execution time penalties of our in-memory
data structure are minimal.


Finally, we note that our proposed techniques are general and can be used for
efficient snapshot retrieval in temporal relational databases as well. In fact, both the DeltaGraph
and the GraphPool data structures treat the network as a collection of objects and do not exploit any
properties of the graphical structure of the data.

\topic{Outline:}  We begin with a discussion of the prior work (Section \ref{sec:related}). We then
discuss the key components of the system, the data model, and present the high level system architecture
(Section \ref{sec:overview}). Then, we describe the DeltaGraph structure in detail (Section~\ref{sec:histstore}), 
and develop analytical models for the storage space and snapshot access times
(Section~\ref{sec:DGanalysis}). We then briefly discuss GraphPool (Section~\ref{sec:inmem}). 
Finally, we present the results of our experimental evaluation
(Section~\ref{sec:experiment}).

\section{Related Work}
\label{sec:related}
There has been an increasing interest in dynamic network analysis over the last decade, fueled by
the increasing availability of large volumes of temporally annotated network data. Many works have
focused on designing analytical models that capture how a network evolves, with a primary focus on
social networks and the
Web (see, e.g., ~\cite{DBLP:series/ads/AggarwalW10a,LeskovecKF07,Kumar:2006:SEO:1150402.1150476}).
There is also much work on understanding how communities evolve, identifying key individuals,
and locating hidden groups in dynamic networks. Berger-Wolf et
al.~\cite{Berger-Wolf2006,Tantipathananandh2007},  
Tang et al.~\cite{Tang2008} and Greene et al.~\cite{Greene2010} 
address the problem of community evolution in
dynamic networks.
McCulloh and Carley~\cite{McCulloh2008} present techniques for social change detection.
Asur et al.~\cite{Asur2009} present a framework for characterizing the complex behavioral patterns
of individuals and communities over time. 
In a recent work, Ahn et al.~\cite{AhnTaxonomy2011} present an exhaustive taxonomy of temporal
visualization tasks. Biologists are interested in discovering historical events leading to a 
known state of a Biological network (e.g.,~\cite{10.1371/journal.pcbi.1001119}).
Ren et al.~\cite{RenEvolvGraph11} analyze evolution of shortest paths
between a pair of vertices over a set of snapshots from the history. 
Our goal in this work is to build a graph data management system that can efficiently and scalably
support these types of dynamic network analysis tasks over large volumes of data in real-time.

There is a vast body of literature on {\em temporal relational databases}, starting with the early
work in the 80's on developing temporal data models and temporal query
languages. We won't attempt to present a exhaustive survey of that work, but instead refer the reader 
to several surveys and books on this
topic~\cite{Bolour92,DBLP:conf/sigmod/SnodgrassA85,Ozsoyoglu1995,Tansel1993,date2002temporal,tsql2,Salzberg1999}.
The most basic concepts that a relational temporal database is based upon are
\textbf{valid time} and \textbf{transaction time}, considered orthogonal to each other. Valid time denotes the time
period during which a fact is true with respect to the real world. Transaction time is the time
when a fact is stored in the database. 
A valid-time temporal database permits correction of a previously incorrectly stored
fact~\cite{DBLP:conf/sigmod/SnodgrassA85}, unlike transaction-time databases where an
inquiry into the past may yield the previously known, perhaps incorrect version of a fact. 
Under this nomenclature, our data management system is based on valid
time -- for us the time the information was entered in the database is not critical, but our focus
is on the time period when the information is true.

From a querying perspective, both valid-time and transaction-time databases can be treated as simply
collections of intervals~\cite{Salzberg1999}, however indexing structures that assume transaction
times can often be simpler since they don't need to support arbitrary inserts or deletes into the
index. Salzberg and Tsotras~\cite{Salzberg1999} present a comprehensive survey of indexing
structures for temporal databases. They also present a classification of different queries that one
may ask over a temporal database. Under their notation, our focus in this work is on the {\bf
valid timeslice} query, where the goal is to retrieve all the entities and their attribute values
that are valid as of a specific time point. We discuss the related work on snapshot retrieval 
queries in more detail in Section \ref{sec:snapshotrelatedwork}.

There has been resurgence of interest in general-purpose graph data management systems 
in both academia and industry. Several commercial and open-source graph management systems are being
actively developed (e.g., Neo4j\footnote{\url{http://www.neo4j.org}}, GBase\footnote{\url{http://www.graphbase.net}}, Pregel~\cite{pregel}, Giraph, Trinity~\cite{trinity},
Cassovary, Pegasus~\cite{Kang:2009:PPG:1674659.1677058}, GPS~\cite{ilprints1039}). 
There is much ongoing work on
efficient techniques for answering various types of queries over graphs and on building indexing
structures for them. However, we are not aware of any graph data management system that focuses
on optimizing snapshot retrieval queries over {\em historical} traces, and on supporting rich temporal 
analysis of large networks.


There is also prior work on temporal RDF data and temporal XML Data. 
Motik~\cite{Motik2010} presents a logic-based approach to representing valid time in RDF and OWL. 
Several works (e.g.,~\cite{Pugliese:2008:SRT:1367497.1367579, Tappolet2009}) have considered the
problems of subgraph pattern matching or SPARQL query evaluation over temporally annotated RDF data.
There is also much work on version management in XML data stores.
Most scientific datasets are semistructured in nature and can be effectively represented in XML
format~\cite{Buneman:2004:ASD:974750.974752}. 
Lam and Wong~\cite{Lam:2003:FIX:1766091.1766101} use \textit{complete deltas}, which can be traversed 
in either direction of time for efficient retrieval. Other systems store the current version as a snapshot
and the historical versions as deltas from the current version~\cite{Marian:2001:CMV:645927.672205}.
For such a system, the deltas only need to be unidirectional. Ghandeharizadeh et
al.~\cite{Ghandeharizadeh:1996:HED:232753.232801} provide a formalism on deltas, which includes a
{\em delta arithmetic}. All these approaches assume unique node identifiers to merge deltas with deltas or
snapshots. Buneman et al.~\cite{Buneman:2004:ASD:974750.974752} propose merging all the versions of the database 
into one single hierarchical data structure for efficient retrieval. 
In a recent work, Seering et al.~\cite{arrayversioning} presented a disk based versioning system using efficient 
delta encoding to minimize space consumption and retrieval time in array-based systems.
However, none of that prior work focuses on snapshot retrieval in general graph databases, or
proposes techniques that can flexibly exploit the memory-resident information.

\section{System Overview}
\label{sec:overview}

We begin with briefly describing our graph data model and categorizing the types of changes that it
may capture. We then discuss different types of snapshot retrieval queries that we 
support in our system, followed by the key components of the system architecture.
 
\subsection{Graph Data Model}
The most basic model of a graph over a period of time is as a collection of {\em graph snapshots}, one
corresponding to each time instance (we assume discrete time). Each such graph snapshot 
contains a set of nodes
and a set of edges. The nodes and edges are assigned unique ids at the time of their creation, which
are not re-assigned after deletion of the components (a deletion followed by a re-insertion results
in assignment of a new id). A node 
or an edge may be associated with a list of attribute-value pairs; the list of attribute names is
not fixed a priori and new attributes may be added at any time. Additionally an edge contains the 
information about whether it is a directed edge or an undirected edge.

We define an \textit{event} as the record of an atomic activity in the network.  An event could
pertain to either the creation or deletion of an edge or node, or change in an attribute value of a
node or an edge. Alternatively, an event can express the occurrence of a  \textit{transient} edge or 
node that is valid only for that time instance instead of an interval (e.g., 
a ``message'' from a node to another node). Being atomic refers to the fact that the activity can not be logically broken down
further into smaller events. Hence, an event always corresponds to a single timepoint. So, the
valid time interval of an edge, $[t_{s}, t_{e}]$, is expressed by two different events, edge
addition and deletion events at $t_{s}$ and $t_{e}$ respectively. 
The exact contents of an event depend on the event type; below we show examples of a new edge event (NE), 
and an update node attribute event (UNA). \\[2pt]
{\tt \scriptsize
(a) \{NE, N:23, N:4590, directed:no, 11/29/03 10:10\} \\
(b) \{UNA, N:23, `job', old:`..', new:`..', 11/29/07 17:00\} \\[2pt]
}
We treat events as bidirectional, i.e., they could be applied to a database
snapshot in either direction of time. For example, say that at times $t_{k-1}$ and $t_{k}$, the graph 
snapshots are $G_{k-1}$ and $G_{k}$ respectively. If $E$ is the set of all events at time $t_{k}$, 
we have that: 
\[ G_{k} = G_{k-1} + E, \ \ \ \ G_{k-1} = G_{k} - E \]

\noindent{where} the $+$ and $-$ operators denote application of the events in $E$ in the forward and the
backward direction. All events are recorded in the direction of evolving time, i.e., going ahead in
time. 
A list of chronologically organized events is called an \textit{eventlist}.

\begin{figure}[t]
  \centering
  \includegraphics[width=0.48\textwidth]{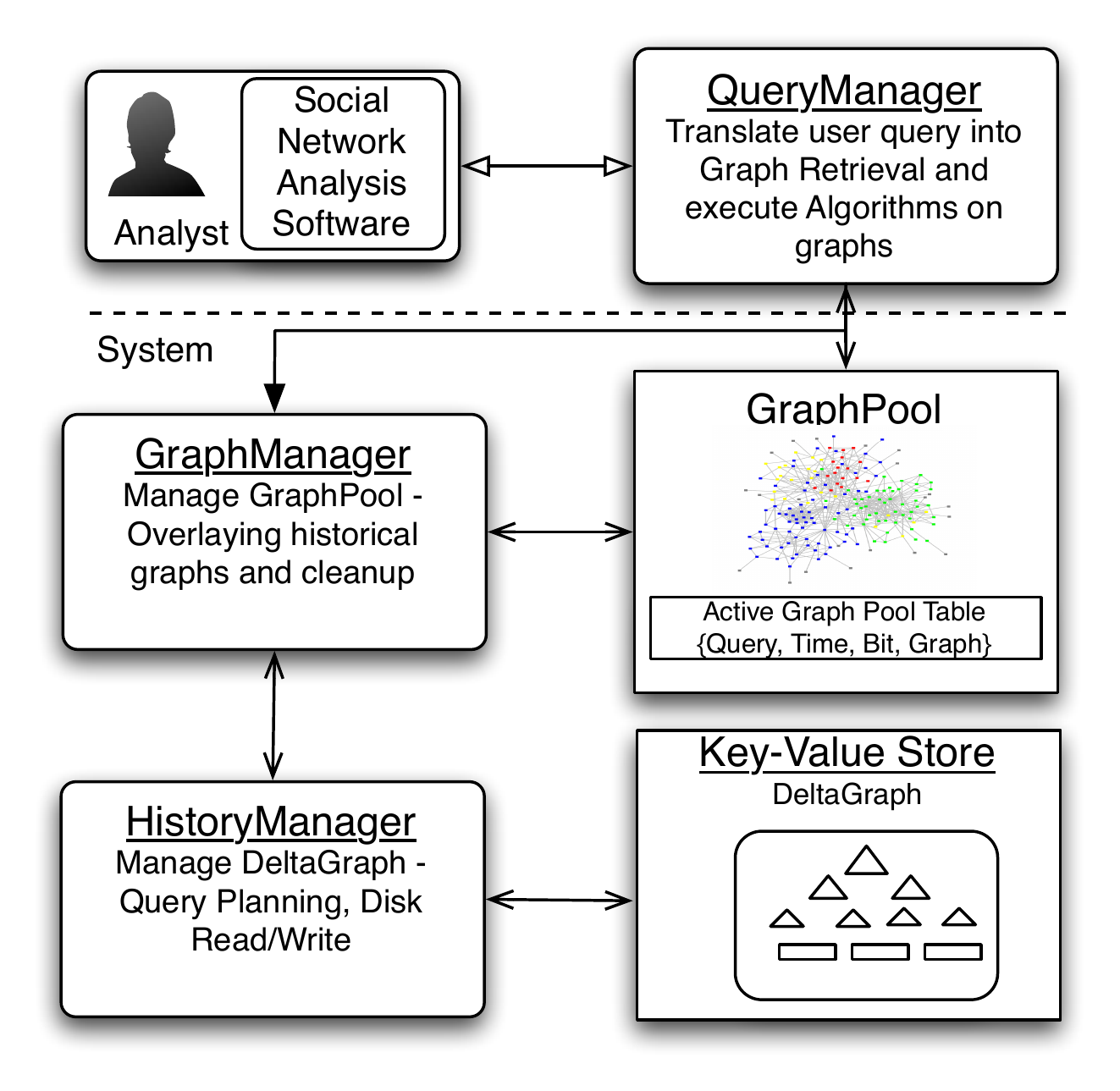}\\
  \caption{Architecture of our system: our focus in this work is on the components below the dashed line.}
  \label{fig:sysarch}
\end{figure}

\subsection{System Overview}
Figure \ref{fig:sysarch} shows a high level overview of our system and its key components. At a
high level, there are multiple ways that a user or an application may interact with a
historical graph database.  Given the wide variety of network analysis or visualization tasks that
are commonly executed against an information network, we expect a large fraction of these
interactions will be through a programmatic API where the user or the application programmer writes
her own code to operate on the graph (as shown in the figure). Such interactions result in what we
call {\em snapshot} queries being executed against the database system. Executing such queries is
the primary focus of this paper, and we further discuss these types of queries below. In ongoing
work, we are also working on developing a high-level declarative query language (similar to
TSQL~\cite{tsql2}) and query processing techniques to execute such queries against our database.
As a concrete example, an analyst who may have designed a new network evolution model
and wants to see how it fits the observed data, may want to retrieve a set of historical snapshots
and process them using the programmatic API. On the other hand, a declarative query
language may better fit the needs of a user interested
in searching for a temporal pattern (e.g., {\em find nodes that had the fastest growth in the number of
neighbors since joining the network}).

Next, we briefly discuss snapshot queries and the key components of the system. 

\subsubsection{Snapshot Queries}
\label{subsec:queries_small}
\label{subsec:queryop}
We differentiate between a {\em singlepoint snapshot query} and a {\em multipoint
snapshot query}. An example of the first query is: ``Retrieve the graph as of January 2, 1995''. 
On the other hand, a multipoint snapshot query requires us to simultaneously retrieve multiple
historical snapshots (e.g, ``Retrieve the graphs as of every Sunday
between 1994 to 2004''). We also support more complex snapshot queries where a {\em TimeExpression}
or a {\em time interval} is specified instead. Any snapshot query can specify whether it requires
only the structure of the graph, or a specified subset of the node or edge attributes, or all
attributes. 

Specifically, the following is a list of some of the retrieval functions that we support in our
programmatic API.
\begin{descriptionsmallermarginnobfnoem}
    \item[{\bf GetHistGraph(Time t, String attr\_options)}:] In the basic singlepoint graph retrieval
        call, the first parameter indicates the time; the second parameter indicates the attribute 
        information to be fetched as a
        string formed by concatenating sub-options listed in Table~\ref{tab:subopts}. For 
        example, to specify that all node attributes except {\em salary}, and the edge attribute {\em name} should be fetched, 
        we would use: attr\_options = ``+node:all-node:salary+edge:name''. 
    \item[{\bf GetHistGraphs(List$<$Time$>$ t\_list, String attr\_options)},] where \\ t\_list specifies a list
        of time points.
    \item[{\bf GetHistGraph(TimeExpression tex, String attr\_options)}:] This is used to retrieve a
        hypothetical graph using a multinomial Boolean expression over time points. For example,
        the expression $(t_{1} \wedge \neg{t_{2}})$ specifies the components of
        the graph that were valid at time $t_{1}$ but not at time $t_{2}$. The TimeExpression data
        structure consists of a list of $k$ time points, $\{t_1, t_2, \dots, t_k\}$, and a Boolean expression over them.

    \item[{\bf GetHistGraphInterval(Time $t_{s}$, Time $t_{e}$, String attr\_options)}:] ~ ~\\This is used to retrieve a 
        graph over all the elements that were added during the time interval $[t_{s},t_{e})$. It
        also fetches the transient events, not fetched (by definition) by the above calls.
\end{descriptionsmallermarginnobfnoem}
        
\begin{table} [t]
\centering
\begin{tabular}{| l | p{5cm} |} 
    \hline
Option & Explanation \\ 
    \hline
    -node:all ({\bf default})& None of the node attributes  \\ 
    \hline
+node:all & All node attributes \\ 
 \hline
+node:attr1 &  Node attribute named ``attr1''; overrides ``-node:all'' for that attribute \\ 
 \hline
 -node:attr1 & Node attribute named ``attr1''; overrides ``+node:all''  for that attribute\\ 
 \hline
\end{tabular}
\caption{Options for node attribute retrieval. Similar options exist for edge attribute retrieval.}
\label{tab:subopts}
\end{table}

%


%

\noindent
The (Java) code snippet below shows an example program that retrieves several graphs, and operates
upon them. \\
\fbox{
\begin{minipage}{0.45\textwidth} 
{\small
/* Loading the index */ \\
GraphManager gm = new GraphManager(\dots);\\
gm.loadDeltaGraphIndex(\dots);\\
\dots \\
/* Retrieve the historical graph structure along with node names as of Jan 2, 1985 */ \\
HistGraph h1 = gm.GetHistGraph(``1/2/1985'', ``+node:name''); \\
\dots \\
/* Traversing the graph*/\\
List$<$HistNode$>$ nodes = h1.getNodes();\\
List$<$HistNode$>$ neighborList = nodes.get(0).getNeighbors();\\
HistEdge ed = h1.getEdgeObj(nodes.get(0), neighborList.get(0));\\
\dots \\
/* Retrieve the historical graph structure alone on Jan 2, 1986 and Jan 2, 1987 */ \\
listOfDates.add(``1/2/1986''); \\
listOfDates.add(``1/2/1987''); \\
List$<$HistGraph$>$ h1 = gm.getHistGraphs(listOfDates, ``''); \\
\dots \\

}
\end{minipage}
}

\smallskip
\noindent
Eventually, our goal is to support Blueprints, a collection of interfaces
analogous to JDBC but for graph data (we currently support a subset). Blueprints is a generic graph Java API that already binds to
various graph database backends (e.g., Neo4j), and many graph processing and
programming frameworks are built on top of it (e.g., Gremlin, a graph traversal
language\footnote{\url{http://github.com/tinkerpop/gremlin/wiki}}; Furnace, a graph algorithms
package\footnote{\url{http://github.com/tinkerpop/furnace/wiki}}; etc.). By supporting
the Blueprints API, we immediately enable use of many of these already existing toolkits.

\subsubsection{Key Components}
\noindent
There are two key data structure components of our system.
\begin{myenumerate}
\item  \textbf{GraphPool} is an in-memory data structure that can store multiple graphs together in
    a compact way by overlaying the graphs on top of each other. At any time, the GraphPool
    contains: (1) the \textit{current graph} that reflects the current state of the network, (2) the
    \textit{historical snapshots}, retrieved from the past
    using the commands above and possibly modified by an application program, and (3) \textit{materialized
    graphs}, which are graphs that correspond interior or leaf nodes in the DeltaGraph, but may not 
    correspond to any valid graph snapshot (Section~\ref{subsec:memmat}). 
    GraphPool
    exploits redundancy amongst the different graph snapshots that need to be retrieved, and considerably 
    reduces the memory requirements for historical queries. 
    We discuss GraphPool in detail in Section \ref{sec:inmem}.

\item  \textbf{DeltaGraph} is an index structure that stores the historical network data using a
    hierarchical index structure over {\em deltas} and {\em leaf-level eventlists} (called {\em
    leaf-eventlists}). 
    To execute a snapshot retrieval query, 
    a set of appropriate deltas and leaf-eventlists is fetched and the resulting graph
    snapshot is overlaid on the existing set of graphs in the GraphPool. The structure of the
    DeltaGraph itself, called {\bf DeltaGraph skeleton}, is maintained as a weighted graph in memory
    (it contains statistics about the deltas and eventlists, but not the actual data). The skeleton is used during query planning
    to choose the optimal set of deltas and eventlists for a given query. We describe DeltaGraph
    in detail in the next section. 
\end{myenumerate}

\noindent{The} data structures are managed and maintained by several system components.
\textit{HistoryManager} deals with the construction of the DeltaGraph, plans how to execute a
singlepoint or a multipoint snapshot query, and reads the required deltas and eventlists from the disk. 
\textit{GraphManager} is responsible for managing the GraphPool data structure, including the overlaying 
of deltas and eventlists, bit assignment, and post-query clean up. Finally, the \textit{QueryManager} 
manages the interface with the user or the application program, and extracts a snapshot query to be executed
against the DeltaGraph. One of its functions is to translate any explicit references (e.g. user-id) from the query to the 
corresponding internal-id and vice-versa for the final result, using a lookup table. As discussed earlier, such 
a component is usually highly application-specific, and we do not discuss it further in this paper. 

In a distributed deployment, DeltaGraph and GraphPool are both partitioned across a set of machines 
by partitioning the node ID space, and assigning each partition to a separate machine (Section~\ref{subsec:dgcons}). 
The partitioning used for storage can be different from that used for retrieval and processing; however, 
for minimizing wasted network communication, it would be ideal for the two partitionings to be aligned so that
multiple DeltaGraph partitions may correspond to a single GraphPool partition, but not vice versa.
Snapshot retrieval on each machine is independent of the others,
and requires no network communication among those. Once the snapshots are loaded into the GraphPool, any distributed
programming framework can be used on top; we have implemented an 
iterative vertex-based message-passing system analogous to Pregel~\cite{pregel}.

For clarity, we assume a single-site deployment (i.e., no horizontal partitioning) in most of the description that follows.


\begin{figure}[t]
\begin{center}
\subfloat{ \includegraphics[width=0.40\textwidth]{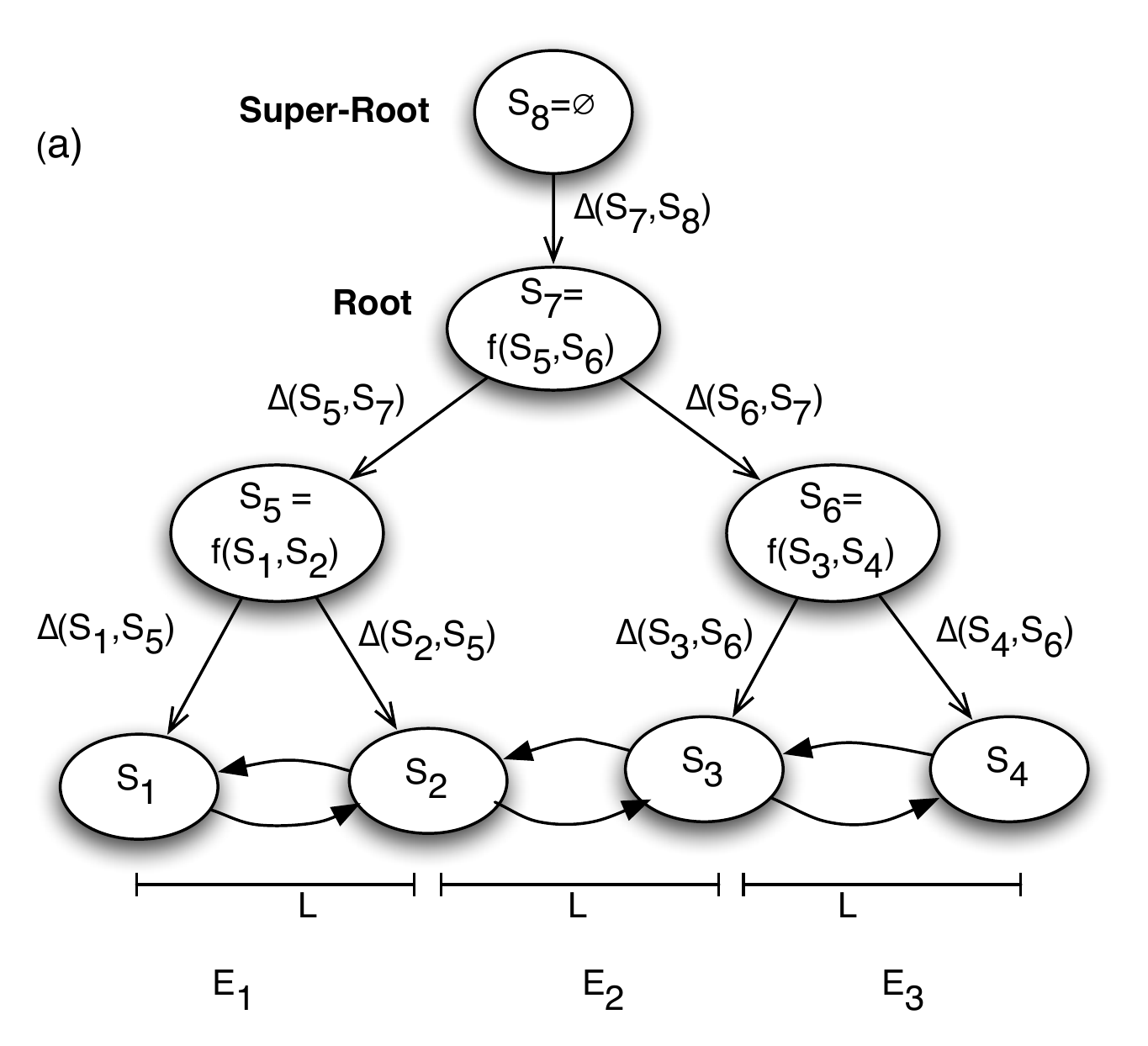}\label{fig:heirsnap}}\\
\subfloat{ \includegraphics[width=0.47\textwidth]{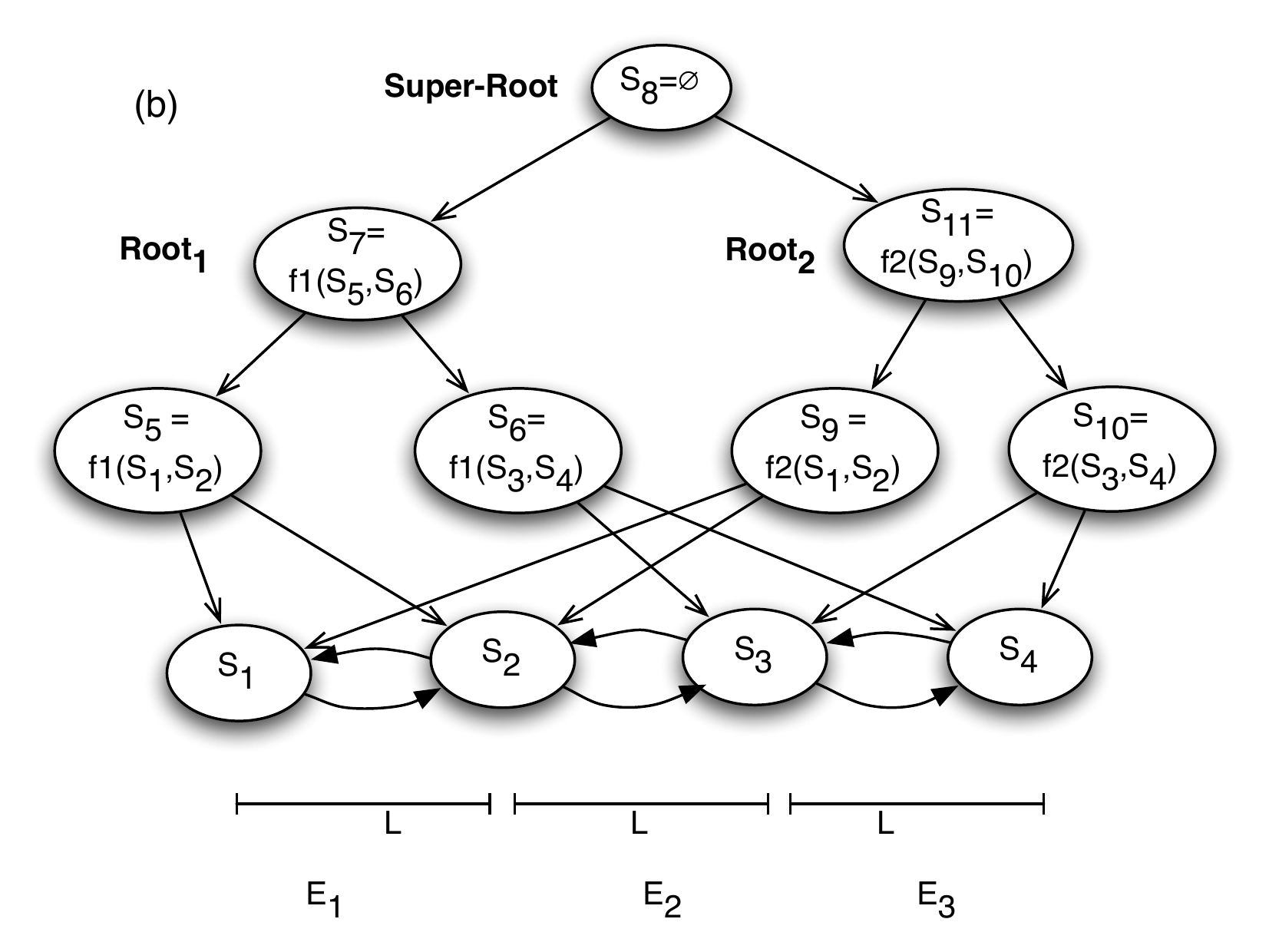}\label{fig:heirsnaphybrid}}
  \caption{DeltaGraphs with $4$ leaves, leaf-eventlist size $L$, arity $2$. $\Delta(S_i,
  S_j)$ denotes delta needed to construct $S_i$ from $S_j$.}
  \label{fig:deltagraphs}
\end{center}
\end{figure}
\section{DeltaGraph: Physical Storage of Historical Graph Data}
\label{sec:histstore}
We begin with discussing previously proposed techniques for supporting snapshot queries, and why
they do not meet our needs. We then present details of our proposed DeltaGraph data structure.

\subsection{Prior Techniques and Limitations}
\label{sec:snapshotrelatedwork}
An optimal solution to answering snapshot retrieval queries is the {\bf external interval tree},
presented by Arge and Vitter~\cite{Arge1996}. Their proposed index structure uses optimal space on
disk, and supports updates in optimal (logarithmic) time.  \textbf{Segment trees}~\cite{Blankenagel1994} can
also be used to solve this problem, but may store some intervals in a duplicated manner and hence
use more space. Tsotras and Kangelaris~\cite{Tsotras1995} present {\bf snapshot index}, an I/O
optimal solution to the problem for transaction-time databases.  Salzberg and
Tsotras~\cite{Salzberg1999} also discuss two extreme approaches to supporting snapshot retrieval
queries, called \textbf{Copy} and \textbf{Log} approaches. In the Copy approach, a snapshot of the
database is stored at each transaction state, the primary benefit being fast retrieval times;
however the space requirements make this approach infeasible in practice.  The other extreme
approach is the Log approach, where only and all the changes are recorded to the database, annotated
by time. While this approach is space-optimal and supports $O(1)$-time updates (for transaction-time
databases), answering a query may require scanning the entire list of changes and takes prohibitive
amount of time.  A mix of those two approaches, called \textbf{Copy+Log}, where a subset of the
snapshots are explicitly stored, is often a better idea.

We found these (and other prior) approaches to be insufficient and inflexible for our needs for
several reasons. First,
they do not efficiently support multipoint queries that we expect to be very commonly used in
evolutionary analysis, that need to be optimized by avoiding duplicate reads and repeated processing
of the events. Second, to cater to the needs of a variety of different applications, we need the index
structure to be highly tunable, and to allow trading off different resources and user
requirements (including memory, disk usage, and query latencies). Ideally we would also like to
control the distribution of average snapshot retrieval times over the history, i.e., we should be
able to reduce the retrieval times for more recent snapshots at the expense of increasing it for the
older snapshots (while keeping the utilization of the other resources the same), or vice-versa.
For achieving low latencies, the index structure should support flexible pre-fetching of 
portions of the index into memory and should avoid processing any events that are not needed by
the query (e.g., if only the network structure is needed, then we should not have to process any 
events pertaining to the node or edge attributes). Finally, we would like the index structure to be
able to support different persistent storage options, ranging from a hard disk to the cloud; most of
the previously proposed index structures are optimized primarily for disks.

\subsection{DeltaGraph Overview}
\label{subsec:deltagraph}
Our proposed index data structure, DeltaGraph, is a directed graphical structure that is largely
hierarchical, with the lowest level of the structure corresponding to equi-spaced historical snapshots 
of the network (equal spacing is not a requirement, but simplifies analysis). Figure \ref{fig:deltagraphs}(a) shows a simple DeltaGraph, where the nodes $S_1,
\dots, S_4$ correspond to four historical snapshots of the graph, spaced $L$ events apart.
We call these nodes {\em leaves}, even though there are bidirectional edges between these nodes as
shown in the figure. 
The interior nodes of the DeltaGraph correspond to graphs that are constructed from its children
by applying what we call a {\bf differential function}, denoted $f()$. For an interior node
$S_p$
with children $S_{c_1}, \dots, S_{c_k}$,\footnote{We abuse the notation somewhat to let $S_p$ denote both the
interior node and the graph corresponding to it.} we have that $S_p = f(S_{c_1}, \dots,
S_{c_k})$. The simplest differential function is perhaps the {\bf Intersection}
function. 
We discuss other differential functions in the next section. 

The graphs $S_p$ are not explicitly stored in the DeltaGraph.
Rather we only store the {\em delta} information with the edges. Specifically, the directed edge from $S_p$ to
$S_{c_i}$ is associated with a delta $\Delta(S_{c_i}, S_p)$ which allows construction of $S_{c_i}$ from
$S_p$. It contains the elements that should be deleted from $S_p$ (i.e., $S_p - S_{c_i}$) and 
those that should be added to $S_p$ (i.e., $S_{c_i} - S_p$).  The bidirectional edges among the
leaves also store similar deltas; here the deltas are simply the eventlists (denoted $E_1, E_2,
E_3$ in Figure \ref{fig:deltagraphs}), called {\bf leaf-eventlists}. For a leaf-eventlist $E$, we denote by $[E^{start}, E^{end})$
the time interval that it corresponds to.
For convenience, we add a special root node, called {\bf super-root}, at the top of the DeltaGraph that is associated with a
null graph ($S_8$ in Figure \ref{fig:deltagraphs}). For convenience, we call the children of the
super-root as {\bf roots}.

A DeltaGraph can simultaneously have multiple hierarchies that use different differential functions
(Figure \ref{fig:deltagraphs}(b)); this can used to improve query latencies at the
expense of higher space requirement.

The deltas and the leaf-eventlists are given unique ids in the DeltaGraph structure, and are stored in a
columnar fashion, by separating out the structure information from the attribute information. For
simplicity, we assume here a separation of a delta $\Delta$ into three components: (1)
$\Delta_{struct}$, (2) $\Delta_{nodeattr}$, and (3) $\Delta_{edgeattr}$. For a leaf-eventlist $E$, we have 
an additional component, $E_{transient}$, where the transient events are stored.

Finally, the deltas and the eventlists are stored in a persistent, distributed key-value store, the key being $\langle partition\_id, delta\_id, c \rangle$, where $c \in \{\delta_{struct},  \delta_{nodeattr}, \dots, E_{transient}\}$ specifies which of the components is being fetched or stored, and $partition\_id$ specifies the partition. Each delta or eventlist is 
typically partitioned and stored in a distributed manner. Based on the $node\_id$ of the concerned node(s), each event, edge, node and attribute is designated a partition, using a hash function $h_p$, such that, $partition\_id = h_p (node\_id)$. In a set up with $k$ distributed storage units, all the deltas and eventlists are likely to have $k$ partitions each. 


\subsection{Singlepoint Snapshot Queries}
\label{subsec:graphretsinglept}
\begin{figure}[t]
\centering
 \subfloat[Singlepoint query $t_{1}$]{ \label{fig:qp1}\includegraphics[width=0.25\textwidth] {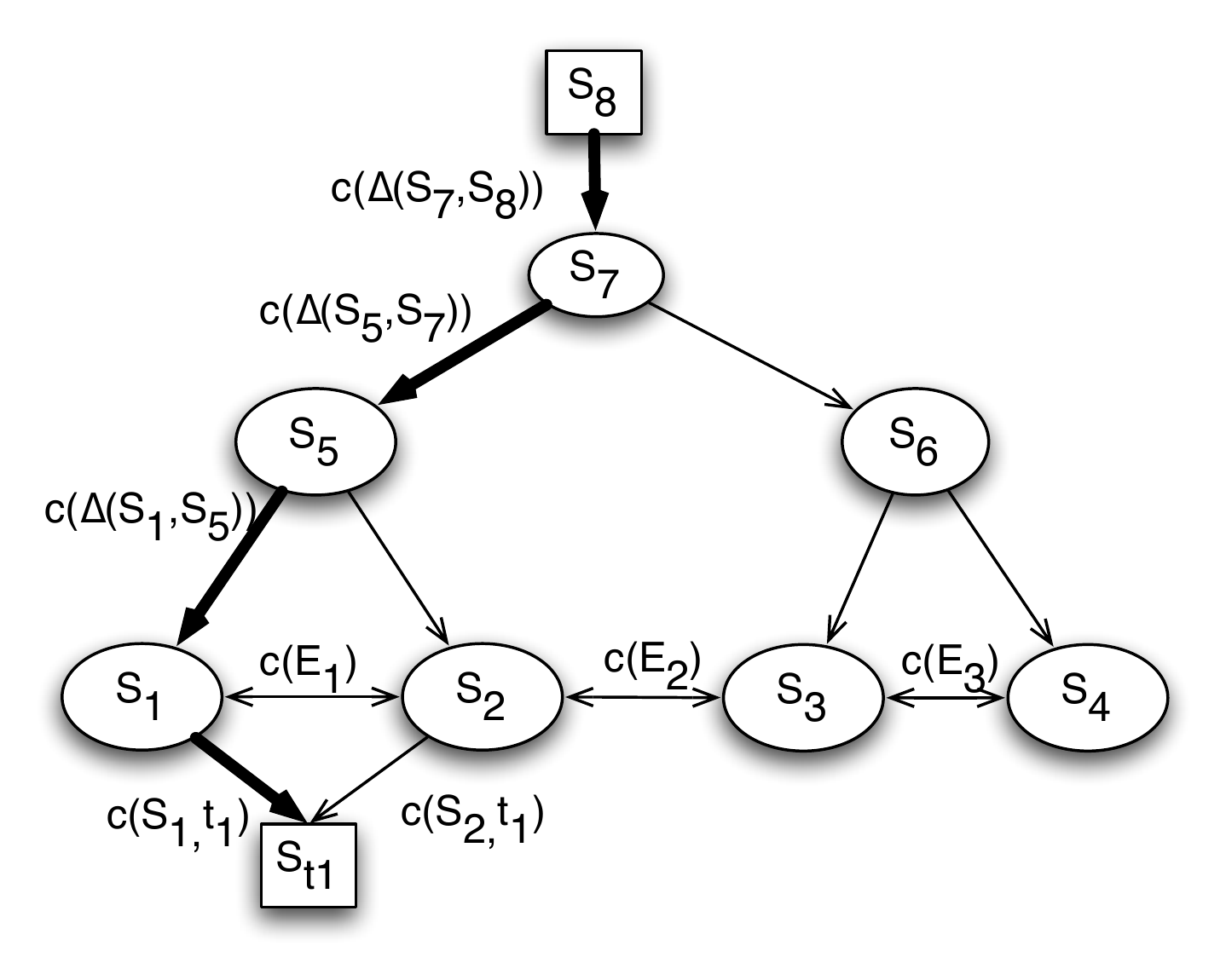}}
 \subfloat[Multipoint query $\{t_{1}, t_{2}, t_{3}\}$]{
 \label{fig:qp2}\includegraphics[width=0.25\textwidth] {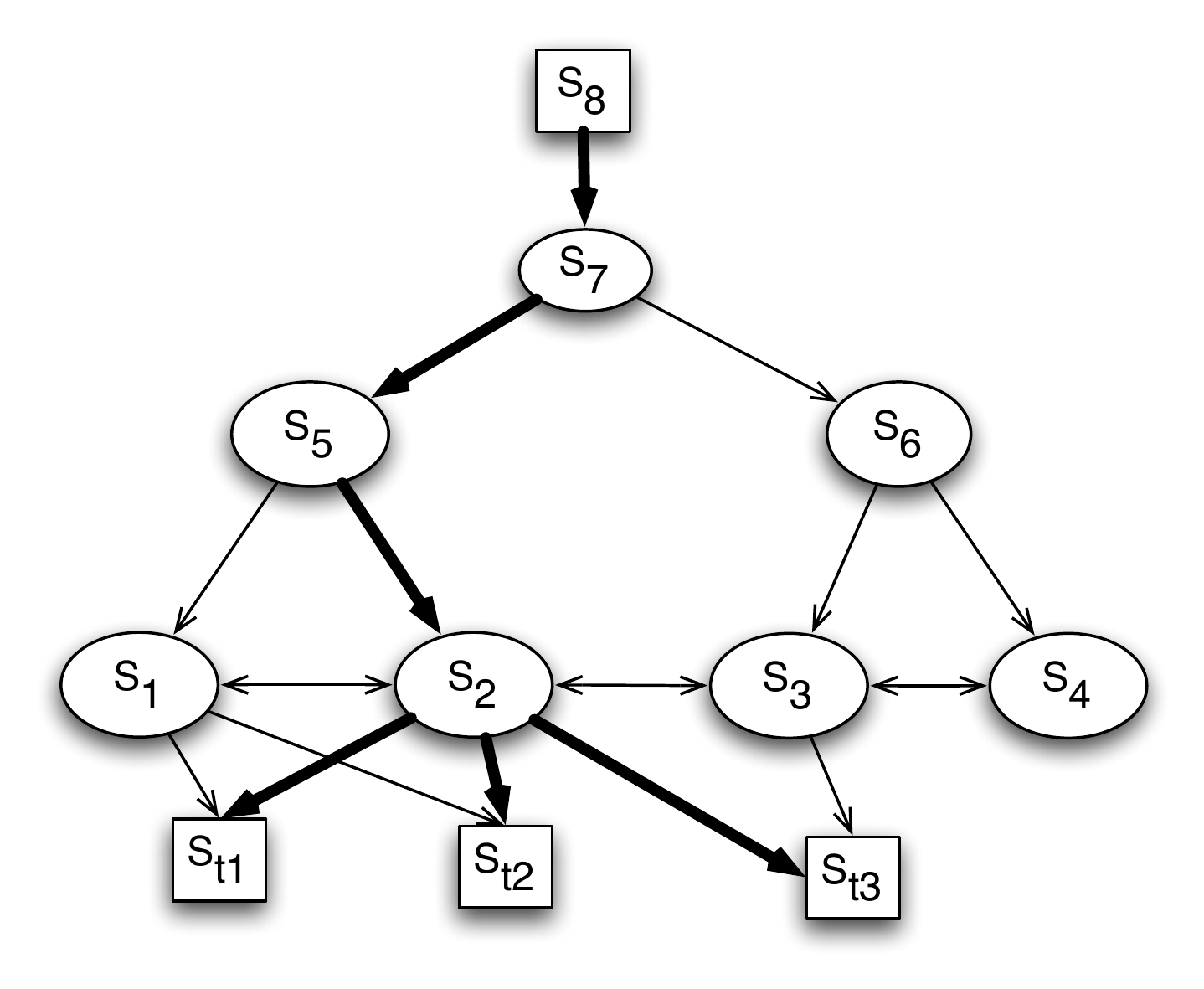}}
  \caption{Example plans for singlepoint and multipoint retrieval on the DeltaGraph shown in Figure 3(a).}
  \label{fig:qplans}
\end{figure}

Given a singlepoint snapshot query at time $t_1$, there are many ways to answer it from the DeltaGraph. 
Let $E$ denote the leaf-eventlist such that $t_1 \in [E^{start}, E^{end})$ (found
through a {\em binary search} at the leaf level). Any
(directed) path from the super-root to the two leaves adjacent to $E$ represents a valid solution to the
query. 
Hence we can find the optimal solution by finding the path with the lowest weight, where the
weight of an edge captures the cost of reading the associated delta (or the required subset of it), and 
applying it to the graph constructed so far. We approximate this cost by using the size of the delta
retrieved as the weight. Note that, each edge is associated with three or four weights, corresponding to
different attr\_options. In the distributed case, we have a set of weights for each partition.
We also add a new virtual node (node $S_{t_1}$ in Figure \ref{fig:qplans}(a)), and 
add edges to it from the adjacent leaves as shown in the figure. The weights associated with these
two edges are set by estimating the portion of the leaf-eventlist $E$ that must be
processed to construct $S_{t_1}$ from those leaves.

We use the standard Dijkstra's shortest path algorithm to find the optimal solution for a specific
singlepoint query, using the appropriate weights. Although this algorithm requires us to traverse
the entire DeltaGraph skeleton for every query, it is needed to handle the continuously changing
DeltaGraph skeleton, especially in response to {\bf memory materialization} (discussed below).
Second, the weights associated with the edges are different for different queries and the
weights are also highly skewed, so the shortest paths can be quite different for the same timepoint
for different attr\_options.
Further, the sizes of the DeltaGraph skeletons (i.e., the number of nodes and edges) are usually small, 
even for very large historical traces, and the running time of the shortest path algorithm is
dwarfed by the cost of retrieving the deltas from the persistent storage. In ongoing work, we are
working on developing an algorithm based on incrementally maintaining single source shortest paths
to handle very large DeltaGraph skeletons.

\subsection{Multipoint Snapshot Queries}
\label{subsec:graphretmultipt}
Similarly to singlepoint snapshot queries, a multipoint snapshot query can be reduced to finding a
{\em Steiner tree} in a weighted directed graph. We illustrate this through an example. Consider a 
multipoint query over three timepoints $t_1, t_2, t_3$ over the DeltaGraph shown in Figure
\ref{fig:deltagraphs}(a). We first identify the leaf-level eventlists that contain the three time
points, and add virtual nodes $S_{t1}, S_{t2}, S_{t3}$, shown in the Figure~\ref{fig:qplans}(b) using shaded nodes. The
optimal solution to construct all three snapshots is then given by the lowest-weight Steiner tree
that connects the super-root and the three virtual nodes (using appropriate weights depending on the
attributes that need to be fetched). A possible Steiner tree is depicted in the figure using
thicker edges. As we can see, the optimal solution to the multipoint query does not use the optimal
solutions for each of the constituent singlepoint queries. Finding the lowest weight Steiner tree is
unfortunately NP-Hard (and much harder for directed graphs vs undirected graphs), and we instead use 
the standard 2-approximation for undirected Steiner trees for that purpose. We first construct a complete undirected 
graph over the set of nodes comprising the root and the virtual nodes, with the weight of an edge between two nodes set to 
be the weight of the shortest path between them in the skeleton. We then compute the minimum spanning tree over this graph, and ``unfold'' 
it to get a Steiner tree over the original skeleton. This algorithm does not work for general directed graphs, 
however we can show that, because of the special structure of a DeltaGraph, it not only results in valid Steiner trees, but retains the 2-approximation guarantee as well. 

Aside from multipoint snapshot queries, this technique is also used
for queries asking for a graph valid for a composite TimeExpression, which we currently execute by
fetching the required snapshots into memory and then operating upon them to find the components that
satisfy the TimeExpression. 

\label{subsec:graphretinterval}

\subsection{Memory Materialization}
\label{subsec:memmat}
For improving query latencies, some nodes in the DeltaGraph are typically pre-fetched and materialized in
memory. In particular, the highest levels of the DeltaGraph should be materialized,
and further, the ``rightmost'' leaf (that corresponds to the current graph) should also be
considered as materialized. The task of materializing one or more DeltaGraph nodes is equivalent to
running a singlepoint or a multipoint snapshot retrieval query, and we can use the algorithms
discussed above for that purpose. After a node is materialized, we modify the in-memory
DeltaGraph skeleton, by adding a directed edge with weight 0 from the super-root to that node. Any
further snapshot retrieval queries will automatically benefit from the materialization.

The option of memory materialization enables fine-grained runtime control over the query latencies and the
memory consumption, without the need to reconstruct the DeltaGraph. For instance, if we know that
a specific analysis task may need access snapshots from a specific period, then we can materialize the
lowest common ancestor of the snapshots from that period to reduce the query latencies. One
extreme case is what we call {\bf total materialization}, where all the leaves are
materialized in memory. This reduces to the Copy+Log approach with the difference that the snapshots
are stored in memory in an overlaid fashion (in the GraphPool). For mostly-growing networks
(that see few deletions), such materialization can be done cheaply resulting in very low
query latencies.


\subsection{DeltaGraph Construction} 
\label{subsec:dgcons}
Besides the graph itself, represented as a list of all events in a chronological order, $E$, the DeltaGraph construction algorithm accepts four other parameters: (1) $L$, the
size of a leaf-level eventlist; (2) $k$, the arity of the graph; (3) $f()$, the differential function that
computes a combined delta from a given set of deltas; and (4) a partitioning of the node ID space. 
The DeltaGraph is constructed in a bottom-up fashion, similar to how a {\em bulkloaded B+-Tree} is
constructed. We scan $E$ from the beginning, creating the leaf snapshots and
corresponding eventlists (containing $L$ events each). When $k$ of the snapshots are created, a parent interior node is
constructed from those snapshots. Then the deltas corresponding to the edges are created, those
snapshots are deleted, and we continue scanning the eventlist.

The entire DeltaGraph can thus be constructed in a single pass over $E$, assuming sufficient memory is
available.  At any point during the
construction, we may have up to $k-1$ snapshots for each level of the DeltaGraph constructed so far.
For higher values of $k$, this can lead to very high memory requirements. However, we use the
GraphPool data structure to maintain these snapshots in an overlaid fashion to decrease the 
total memory consumption. We were able to scale to reasonably large graphs using this technique. 
Further scalability is achieved by making multiple passes over $E$, processing one partition (Section \ref{subsec:deltagraph}) in each pass. 

\subsection{Extensibility}
\label{sec:extensible}
To efficiently support specific types of queries or tasks, it is beneficial to maintain and index
auxiliary information in the DeltaGraph and use it during query execution. 
We extend the DeltaGraph functionality through user-defined modules and functions for this purpose. 
In essence, the user can supply 
functions that compute auxiliary information for each snapshot, that will be automatically indexed along with
the original graph data. The user may also supply a different differential function to be used for
this auxiliary information. 
The basic retrieval functionality (i.e., retrieve snapshots as of a specified time points) is thus naturally extended
to such auxiliary information, which can also be loaded into memory and operated upon. In addition, the user may also 
supply functions that operate on the auxiliary information deltas during retrieval, that can be used to directly 
answer specific types of queries. 

The extensibility framework involves the \textit{AuxiliarySnapshot} and \textit{AuxiliaryEvent} 
structures which are similar to the graph snapshot and event structures, respectively. An AuxiliarySnapshot consists of a hashtable of
string key-value pairs where as an AuxiliaryEvent consists of the event timestamp, a flag indicating the addition, deletion or the change of  a
key-value pair, and finally, a key-value pair itself. Given the very general nature of the auxiliary structures, it is possible for the consumer 
(programs using this API) to define a wide variety of graph indexing semantics whose historical indexing will be done automatically by the HistoryManager. 
For a historical graph, any number of auxiliary indexes may be used, each one by extending the \textit{AuxIndex} abstract class, defining the following, a) method \textit{CreateAuxEvent} that generates an AuxiliaryEvent corresponding to a plain Event, based upon the current Graph and the latest Auxiliary Snapshot,  b) method \textit{CreateAuxSnapshot}, to create a leaf-level AuxiliarySnapshot based upon the previous AuxilarySnapshot and the AuxiliaryEventList in between the two, and c) a method \textit{AuxDF}, a differential function that computes the parent AuxiliarySnapshot, given a list of $k$ AuxiliarySnapshots and corresponding $k-1$ AuxEventlists. 

Any number of queries may be defined on an auxiliary index by extending either of the three \textit{Auxiliary Query} abstract classes, namely, \textit{AuxHistQueryPoint}, \textit{AuxHistQueryInterval} or \textit{AuxHistQuery}, depending on the temporal nature of the query - point, range or across entire time span, respectively. An implementation of any of these classes involves attaching a pointer to the AuxiliaryIndex which the queries are supposed to operate upon (in addition to the plain DeltaGraph index itself), and the particular query method, \textit{AuxQuery}. While doing so, the the programmer typically makes use of the methods like \textit{GetAuxSnapshot} provided by the Extensibility API.

%
%

We illustrate this extensibility through an example of a {\em subgraph pattern matching} index.
 Techniques for subgraph pattern matching are very well studied in literature (see, e.g.,~\cite{he:sigmod08}).
Say we want to support finding all instances of a node-labeled query graph (called {\em pattern graph}) 
in a node-labeled data graph. One simple way to efficiently support such queries is to index all paths of say length 4 in the data graph~\cite{Shasha:2002:AAT:543613.543620}. 
This pattern index here takes the form of a {\em key-value} data structure, where a {\em key} is a quartet of labels, and 
the value is the set of all paths in the data graph over 4 nodes that match it. To find all matching instances for
a pattern, we then decompose the pattern into paths of length 4 (there must be at least one such path in the pattern), use the 
index to find
the sets of paths that match those decomposed paths, and do an appropriate join to construct the entire match. 

We can extend the basic DeltaGraph structure to support such an index as follows. The auxiliary information maintained is
precisely the pattern index at each snapshot, which is naturally stored in a compact fashion in the DeltaGraph (by exploiting
the commonalities over time). It involves implementation of the AuxIndex and AuxHistQueryInterval classes accordingly. 
For example, the \textit{CreateAuxEvent}, creates an AuxEvent, defining the addition or deletion of a path 
of four nodes, by finding the effect of a plain Event (in terms of paths) in the context of the current graph. Instead of using the standard differential function, 
we use one that achieves the following effect: a specific path containing 4 nodes, say $\langle n_1, n_2, n_3, n_4 \rangle$, is present in 
the pattern index associated with an interior node if and only if, it is present in all the snapshots below that interior node. 
This means that, if the path is associated with the root, it is present throughout the history of the network. Such 
auxiliary information can now be directly used to answer subgraph pattern matching query to find
all matching instances over the history of the graph. 
We evaluated our implementation on Dataset 1 (details in the Section \ref{sec:experiment}), and assigned labels to each node by
randomly picking one from a list of ten labels. We built the index as described above (by indexing all paths of
length 4). We were able to run a subgraph pattern query in 148 seconds to find all occurrences of a given pattern
query, returning a total of 14109 matches over the entire history of the network.

This extensibility framework enables us to quickly design and deploy reasonable strategies for answering
many different types of queries. We note that, for specific queries, it may be possible to design more efficient strategies.
In ongoing work, we are investigating the issues in answering specific types of queries over historical 
graphs, including subgraph pattern matching, reachability, etc.

\section{DeltaGraph Analysis}
\label{sec:DGanalysis}
\noindent
Next we develop analytical models for the storage requirements, memory
consumption, and query latencies for a DeltaGraph. 

\subsection{Model of Graph Dynamics}
Let $G_0$ denote the initial graph as of time $0$, and let $G_{|E|}$ denote the graph after $|E|$
events. To develop the analytical models, we make some simplifying assumptions, the most critical
being that  we assume a constant rate of inserts or deletes. Specifically, we assume that a $\delta_*$ fraction
of the events result in an addition of an element to a graph (i.e., inserts), and $\rho_*$ fraction of the events
result in removal of an existing element from the graph (deletes). An update is captured as a delete
followed by an insert. Thus, we have that $|G_{|E|}| = |G_0| + |E| \times \delta_* - |E| \times \rho_*$.
We have that $\delta_* + \rho_* < 1$, but not necessarily $= 1$ because of
transient events that don't affect the graph size. 
Typically we have that $\delta_* > \rho_*$. If $\rho_* = 0$, we call the graph a {\em growing-only}
graph.

Note that, the above model does not require that the graph change at a constant rate {\em over time}. 
In fact, the above model (and the DeltaGraph structure) don't explicitly reason about time but
rather only about the events. To reason about graph dynamics over time, we need a model that
captures {\em event density}, i.e., number of events that take place over a period of time. 
Let $g(t)$ denote the total number of events that take place from time $0$ to time $t$. 
For most real-world networks, we expect $g(t)$ to be a super-linear function of
$t$, indicating that the rate of change over time itself increases over time.

\subsection{Differential Functions}
Recall that a differential function specifies how the snapshot corresponding to an interior node
should be constructed from snapshots corresponding to its children. The simplest differential
function is {\em intersection}. However, for most networks, intersection does not lead to
desirable behavior. For a growing-only graph, intersection results in a left-skewed DeltaGraph,
where the delta sizes are lower on the part corresponding to the older snapshots. In fact, the root
is exactly $G_0$ for a strictly growing-only graph. 

Table \ref{tab:difffuns} shows several other differential functions with better and tunable
behavior. Let $p$ be an interior node with children $a$ and $b$. Let $\Delta(a, p)$ and $\Delta(b,
p)$ denote the corresponding deltas. Further, let $b = a + \delta_{ab} - \rho_{ab}$. 
\begin{descriptionsmallermarginnobfnoem}
    \item[{\bf Skewed:}] For the two extreme cases, $r = 0$ and $r = 1$, we have that $f(a, b) = a$ and 
        $f(a, b) = b$ respectively. By using an appropriate value of $r$, we can control the sizes
        of the two deltas. For example, for $r = 0.5$, we get $p = a + \frac{1}{2}
        \delta_{ab}$. Here $\frac{1}{2} \delta_{ab}$ means that we randomly choose half of the events that
        comprise $\delta_{ab}$ (by using a hash function that maps the events to 0 or 1).
        So $|\Delta(a,p)| = \frac{1}{2} |\delta_{ab}|$, and 
        $|\Delta(b,p)| = \frac{1}{2} |\delta_{ab}| + |\rho_{ab}|$. 
    \item[{\bf Balanced:}] This differential function, a special case of {\bf mixed}, ensures that the delta sizes are balanced
        across $a$ and $b$, i.e., $|\Delta(a, p)| = |\Delta(b, p)| = \frac{1}{2}|\delta_{ab}| +
        \frac{1}{2}|\rho_{ab}|$. Note that, here we make an assumption that $a + \frac{1}{2}\delta_{ab} -
        \frac{1}{2}\rho_{ab}$ is a valid operation. A problem may occur because an event $\in
        \rho_{ab}$ may be selected for removal, but may not exist in $a + \frac{1}{2}\delta_{ab}$.
        We can ensure that this does not happen by using the same hash function for choosing both
        $\frac{1}{2}\delta_{ab}$ and $\frac{1}{2}\rho_{ab}$.
    \item[{\bf Empty:}] This special case makes the DeltaGraph approach identical to the Copy+Log
        approach.
\end{descriptionsmallermarginnobfnoem}
The other functions shown in Table \ref{tab:difffuns} can be used to expose more subtle trade-offs,
but our experience with these functions suggests that, in practice, Intersection, Union, and Mixed functions
are likely to be sufficient for most situations.

\begin{table} [t]
\caption{Differential Functions}
\label{tab:difffuns}
\begin{tabular}{| l | p{6cm}|}
    \hline
  Name & Description \\
    \hline
Intersection & $f(a,b, c  \dots) = a \cap b   \cap c \dots$ \\
\hline
Union & $f(a,b, c  \dots)  = a \cup b \cup c \dots$ \\
 \hline
Skewed & $f(a,b) = a + r.(b-a)$, $0 \le r \le 1$  \\
 \hline
 Right Skewed & $f(a,b) = a \cap b + r.(b - a \cap b)$, $0 \le r \le 1$  \\
  \hline
 Left Skewed & $f(a,b) = a \cap b + r.(a - a \cap b)$, $0 \le r \le 1$  \\
   \hline
Mixed & $f(a,b, c \dots) = a + r_{1}.(\delta_{ab} + \delta_{bc}  \dots) - r_{2}.(\rho_{ab} + \rho_{bc}  \dots)$, $0 \le r_{2} \le r_{1} \le 1$  \\
\hline
Balanced & $f(a,b, c \dots) = a + {1 \over 2}.(\delta_{ab} + \delta_{bc}  \dots) -{1 \over 2}.(\rho_{ab} + \rho_{bc}  \dots)$  \\
  \hline
   Empty & $f(a,b, c \dots) = \emptyset$  \\
  \hline
\end{tabular}
\end{table}

\subsection{Space and Time Estimation Models}
\label{subsec:est_models}
Next, we develop analytical models for various quantities of interest in the DeltaGraph, including
the space required to store it, the distribution of the delta sizes across levels, and the snapshot
retrieval times. We focus on the Balanced and Intersection differential functions, and omit
detailed derivations for lack of space. 

We make several simplifying assumptions in the analysis below. As discussed above, we assume
constant rates of inserts and deletes. Let $L$ denote the leaf-eventlist size,
and let $E$ denote the complete eventlist corresponding to the historical trace. Thus, we have $N =
\frac{|E|}{L}+1$ leaf nodes. We denote by $k$ the arity of the graph, and assume that $N$ is a power
of $k$ (resulting in a complete $k$-ary tree). We number the DeltaGraph levels from the
bottom, starting with 1 (i.e., the bottommost level is called the first level).

\topic{Balanced Function:} Although it appears somewhat complex, the Balanced differential function is
the easiest to analyze. Consider an interior node $p$ with $k$ children, $c_1, \dots, c_k$. If $p$
is at level 2 (i.e., if $c_i$'s are leaves), then $S_p = S_{c_1} + {1 \over 2} \delta_{c_1c_2} -
{1 \over 2}\rho_{c_1c_2} +  {1 \over 2}\delta_{c_2c_3} - \cdots$. It is easy to show that:
 \begin{align*}
 |\Delta(p, c_i)| &= {1 \over 2} (|\delta_{c_1c_2}| + |\rho_{c_1c_2}| + \cdots),\ \ \ \  \forall
 i\\ 
  &= {1 \over 2} (k - 1) (\delta_* + \rho_*) L ,\ \ \ \  \forall i
 \end{align*}
The number of edges between the nodes at the first and second levels is $N$, thus the total space required by the
deltas at this level is: $N {1 \over 2} (k - 1) (\delta_* + \rho_*) L = {1 \over 2} (k - 1)(\delta_* +
\rho_*) |E|$.

If $p$ is an interior node at level 3, then the distance between $c_1$ and $c_2$ in terms of the
number of events is exactly $k (\delta_* + \rho_*) L$. This is because $c_2$ contains: (1) the insert
and delete events from $c_1$'s children that $c_1$ does not contain, (2) $(\delta_* + \rho_*)$ events
that occur between $c_1$'s last child and $c_2$'s first child, and (3) a further ${(k-1) \over 2}
(\delta_* + \rho_*) L$ events from its own children. Using a similar reasoning to above, we can see
that:\\[2pt]
\centerline{$|\Delta(p, c_i)| = {1 \over 2} (k - 1) k (\delta_* + \rho_*) L ,\ \ \ \  \forall i$}
\\[2pt]
Surprisingly, because the number of edges at this level is $N \over k$, the total space occupied
by the deltas at this level is same as that at the first level, and this is true for the higher
levels as well.

Thus, the total amount of space required to store all the deltas (excluding the one from the empty
super-root node to the root) is: \\[2pt]
\centerline{${(\log_k N - 1) \over 2} (k - 1) (\delta_* + \rho_*) |E|$}\\[2pt]
The size of the snapshot corresponding to the root itself can be seen to be: $ |G_0| + {1 \over 2} (\delta_* -
\rho_*) |E|$ (independent of $k$).
Although this may seem high, we note that the size of the current graph ($G_{|E|}$) is:
$ |G_0| + (\delta_* - \rho_*) |E|$, which is larger than the size of the root. 
Further, there is a significant overlap between the two, especially if $|G_0|$ is large, making it 
relatively cheap to materialize the root. 

Finally, using the symmetry, we can show that the total weight of the shortest path between the root 
and any leaf is: ${1 \over 2} (\delta_* + \rho_*) |E|$, resulting in balanced query latencies for
the snapshots (for specific timepoints corresponding to the same leaf-eventlist, there are small
variations because of different portions of the leaf-eventlist that need to be processed).

\topic{Intersection:} On the other hand, the Intersection function is much trickier to analyze. In
fact, just calculating the size of the intersection for a sequence of snapshots is non-trivial in
the general case. As above, consider a graph containing $|E|$ events. The root of the DeltaGraph
contains all events that were not deleted from $G_0$ during that event trace. We state the following
analytical formulas for the size of the root for some special cases without full derivations.
\begin{descriptionsmallermarginnobfnoem}
    \item[\underline{$\rho_* = 0$}:] For a growing-only graph, root snapshot is exactly $G_0$.
    \item[\underline{$\delta_* = \rho_*$}:] In this case, the size of the graph remains constant (i.e., $G_{|E|} =
        G_0$). We can show that: $|root| = |G_0| e^{-{|E|\delta_* \over |G_0|}}$.
    \item[\underline{$\delta_* = 2\rho_*$}:] $|root| = {|G_0|^2 \over |G_0| + \rho_* |E|}$.
\end{descriptionsmallermarginnobfnoem}

The last two formulas both confirm our intuition that, as the total number of events increases, the
size of the root goes to zero.  Similar expressions can be derived for the sizes of any specific
interior node or the deltas, by plugging in appropriate values of $|E|$ and $|G_0|$. We omit
resulting expressions for the total size of the index for the latter two cases.

The Intersection function does have a highly desirable property that, the total weight of the shortest 
path between the super-root and any leaf, is exactly the size of that leaf. Since an interior node
contains a subset of the events in each of its children, we only need to fetch the remaining events
to construct the child. However, this means that the query latencies are typically skewed, with the
older snapshots requiring less time to construct than the newer snapshots (that are typically
larger).

\subsection{Discussion}
\label{subsec:difffunct}
We briefly discuss the impact of different construction parameters and suggest strategies for
choosing the right parameters. We then briefly present a qualitative comparison with interval trees,
segment trees, and the Copy+Log approach.

\topic{Effect of different construction parameters:} 
The parameters involved in the construction of the DeltaGraph give it high flexibility, and must be 
chosen carefully. The optimal choice of the parameters is highly dependent on the application
scenario and requirements. The effect of arity is easy to quantify in most cases: higher arity
results in lower query access times, but usually much higher disk space utilization (even for
the Balanced function, the query access time becomes dependent on $k$ for a more realistic cost model
where using a higher number of queries to fetch the same amount of information takes more time).
Parameters such as $r$ (for Skewed function) and $r_1, r_2$ (for Mixed function) can be used to
control the query retrieval times over the span of the eventlist. For instance, if we expect a
larger number of queries to be accessing newer snapshots, then we should choose higher values for
these parameters. 

The choice of differential function itself is quite critical. Intersection typically leads to lower
disk space utilization, but also highly skewed query latencies that cannot be tuned except through memory
materialization.  Most other differential functions lead to higher disk utilization but provide
better control over the query latencies. Thus if disk utilization is of paramount importance, then
Intersection would be the preferred option, but otherwise, the Mixed function (with the values of
$r_1$ and $r_2$ set according to the expected query workload) would be the recommended option.

Fine-tuning the values of these parameters also requires knowledge of $g(t)$, the event density over
time. The analytical models that we have developed reason about the retrieval times for the leaf
snapshots, but these must be weighted using $g(t)$ to reason about retrieval times over time. 
For example, the Balanced function does not lead to uniform query latencies over time for
graphs that show super-linear growth. Instead, we must choose $r_1, r_2 > 0.5$ to guarantee uniform
query latencies over time in that case.

\topic{Qualitative comparison with other approaches:}
The Copy+Log approach can be seen as a special case of DeltaGraph with Empty
differential function (and arity = $N$). Compared to interval trees, DeltaGraph will almost always
need more space, but its space consumption is usually lower than segment trees. Assume that $\delta_* +
\rho_* = 1$ (this is the worst case for DeltaGraph).
Then, for the Balanced function, with arity ($k$) = 2, the disk space required is
$O(|E|\log N)$. Since the number of {\em intervals} is at least $|E|/2$, the space requirements for
interval trees and segment trees are $O(|E|)$ and $O(|E|\log|E|)$ respectively. For growing-only
graphs and the Intersection function, we see similar behavior. In most other scenarios, we expect
the total space requirement for DeltaGraph to be somewhere in between $O(|E|)$ and $O(|E| \log N)$, and 
lower if $\delta_* + \rho_* \ll 1$ (which is often the case for social networks).

Regarding query latencies, for the Intersection function without any materialization, the amount of
information retrieved for answering a snapshot query is exactly the size of the snapshot. Both
interval trees and segment trees behave similarly. On the other hand, if the root or some of the
higher levels of the DeltaGraph are materialized, then the query latencies could be significantly
lower than what we can achieve with either of those approaches. For Balanced function, if the root
is materialized, then the {\em average} query latencies are similar for the three approaches.
However, for the Balanced function, the retrieval times do not depend on the size of the retrieved
snapshot, unlike interval and segment trees, leading to more predictable and desirable behavior.
Again, with materialization, the query latencies can be brought down even further.

\section{GraphPool}
\label{sec:inmem}
The in-memory graphs are stored in the in-memory \textit{GraphPool} in an overlapping manner.
In this section, we briefly describe the key ideas behind this data structure.

\begin{figure}[t]
\centering
\includegraphics [width=0.5\textwidth]{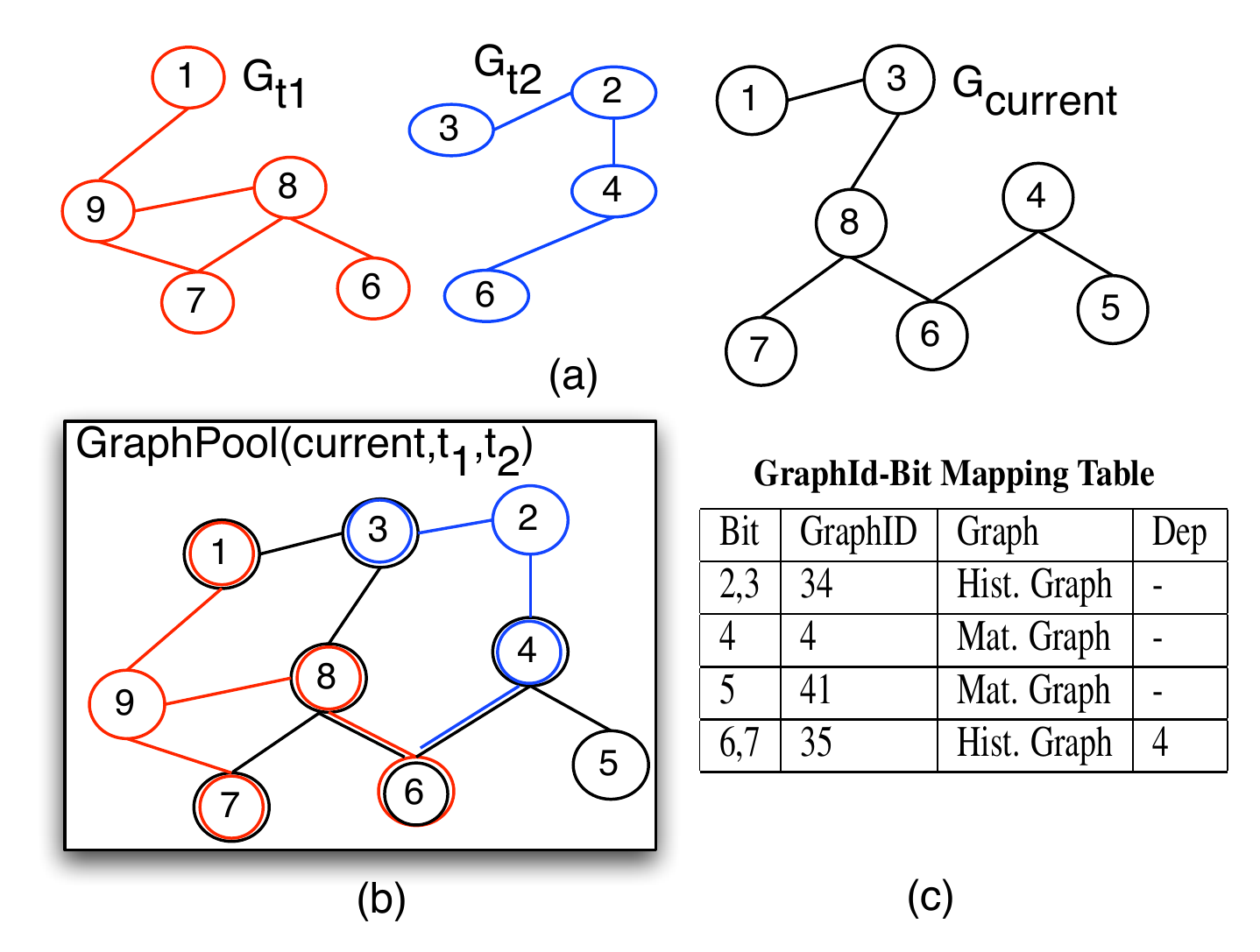}
\caption{(a) Graphs at times $t_{current}$, $t_{1}$ and $t_{2}$; (b) GraphPool consisting of overlaid graphs; (c) GraphID-Bit Mapping Table}\label{fig:gr5}
\end{figure}

\topic{Description:} GraphPool maintains a single graph that is the union of all the \textit{active} graphs including: (1) the
current graph, (2) historical snapshots, and (3) materialized graphs (Figure \ref{fig:gr5}). 
Each component (node or edge), and for each attribute, each of its possible attribute values, are associated with a bitmap string
(called BM henceforth), 
used to decide which of the active graphs contain that component or attribute. A {\em GraphID-Bit} mapping
table is used to maintain the mapping of bits to different graphs. Figure~\ref{fig:gr5}(c) shows an
example of such a mapping. Each historical graph that has been fetched is assigned two consecutive
Bits, $\{2i, 2i+1\}, i \ge 1$. Materialized graphs, on the other hand, are only assigned one Bit. 

Bits 0 and 1
are reserved for the current graph membership. Specifically, Bit 0 tells us whether the element
belongs to the current graph or not. Bit 1, on the other hand, is used for elements that may have
been recently deleted, but are not part of the DeltaGraph index yet.
A Bit associated with a materialized graph is interpreted in a straightforward manner.

Using a single bit for a historical graph misses out on a significant optimization opportunity. Even
if a historical graph differs from the current graph or one of the materialized graphs in only a few
elements, we would still have to set the corresponding bit appropriately for all the elements in the graph. 
We can use the bit pair, $\{2i, 2i+1\}$, to eliminate this step. We mark the historical graph as being dependent on
a materialized graph (or the current graph) in such a case. For example, in Figure~\ref{fig:gr5}(c), historical
snapshot 35 is dependent on materialized graph 4. If Bit $2i$ is true, then the membership of
an element in the historical graph is identical to its membership in the materialized graph (i.e.,
if present in one, then present in another). On the other hand, if Bit $2i$ is false, then 
Bit $2i+1$ tells us whether the element is contained in the historical graph or not (independent
of the materialized graph).

When a graph is pulled into the memory either in response to a query or for materialization, it is overlaid onto 
the current in-memory graph, edge by edge and node by node. The number of graphs that can be
overlaid simultaneously depends primarily on the amount of memory required to contain the union of
all the graphs. The bitmap size is dynamically adjusted to accommodate more graphs if needed, and overall does 
not occupy significant space. The
determination of whether to store a graph as being dependent on a materialized graph is made at the
query time. During the query plan construction, we count the total number of events that need to be
applied to the materialized graph, and if it is small relative to the size of the graph, then the
fetched graph is marked as being dependent on the materialized graph.


\topic{Updates to the Current graph:} As the current graph is being updated, the DeltaGraph index is
continuously updated. All the new events are recorded in a {\em recent} eventlist. When the
eventlist reaches sufficient size (i.e., $L$), the eventlist is inserted into the index and the
index is updated by adding appropriate edges and deltas. We omit further details because of lack of space.

\topic{Clean-up of a graph from memory:} When a historical graph is no longer needed, it needs to be
\textit{cleaned}. Cleaning up a graph is logically a reverse process of fetching it into the
memory. The naive way would be to go through all
the elements in the graph, and reset the appropriate bit(s), and delete the element if no bits are
set. 
However the cost of doing this can be quite high. We instead perform clean-up in a lazy fashion,
periodically scanning the GraphPool in the absence of query load, to reset the bits, and to see if any elements should be deleted. 
Also, in case the system is running low on memory and there are one or more unneeded graphs, the Cleaner thread is invoked and not interrupted until the desired amount of memory is liberated.

\section{Empirical Evaluation}
\label{sec:experiment}

In this section, we present the results of a comprehensive experimental evaluation conducted to
evaluate the performance of our prototype system, implemented in Java using the Kyoto Cabinet key-value
store as the underlying persistent storage.
The system provides a programmatic API including the API discussed in
Section \ref{subsec:queries_small}; in addition, we have implemented a Pregel-like iterative framework
for distributed processing, and the subgraph pattern matching index presented in Section \ref{sec:extensible}.

\topic{Datasets:} We used three datasets in our experimental study. \\[2pt]
\textbf{(1) Dataset 1} is a growing-only, co-authorship network extracted from the DBLP dataset, 
with $2M$ edges in all. 
The network starts empty and grows over a period of seven decades. The nodes (authors) and edges (co-author relationships) are added to the network, 
and no nodes or edges are dropped. At the end, the total number of unique nodes present
in the graph is around 330,000, and the number of edges with unique end points is $1.04M$. Each node was assigned 10 attribute key-value pairs, generated at
random.
\\[2pt]
\textbf{(2) Dataset 2} is a randomly generated historical trace with Dataset 1 as the starting snapshot, followed by $2M$ events where $1M$
edges added and $1M$ edges are deleted over time. 
\\[2pt]
\textbf{(3) Dataset 3} is a randomly generated historical trace with a starting snapshot containing  $10$ million ($10M$) edges and $3M$ nodes (from a patent
        citation network), followed by $100M$ events, $50M$ edge additions and $50M$ edge deletions. 




\topic{Experimental Setup:} 
We created a partitioned index for Dataset 3 and deployed a parallel framework for PageRank
computation using 7 machines, each with a single Amazon EC2 core and approximately 1.4GB of available memory.
Each DeltaGraph partition was approximately 2.2GB. Note that the index is stored in a compressed fashion (using
built-in compression in Kyoto Cabinet). On average, it took us 23.8 seconds to calculate PageRank for a 
specific graph snapshot, including the snapshot retrieval time. This experiment illustrates the effectiveness
of our framework at scalably handling large historical graphs.

For the rest of the experimental study, we report results for Datasets 1 and 2;
the techniques we compare against are centralized, and further the cost of constructing
the index makes it hard to run experiments that evaluate the effect of the construction parameters.
Unless otherwise specified, the experiments were run on a single Amazon EC2 core (with 1.4GB memory).

\topic{Comparison with other storage approaches:}
We begin with comparing our approach with {\bf in-memory} interval trees, and Copy+Log approach.
Both of those were integrated into our system such that any of the approaches could be used to fetch
the historical snapshots into the GraphPool, and we report the time taken to do so. 

\begin{figure}[t]
\centering
 \includegraphics [width=0.5\textwidth] {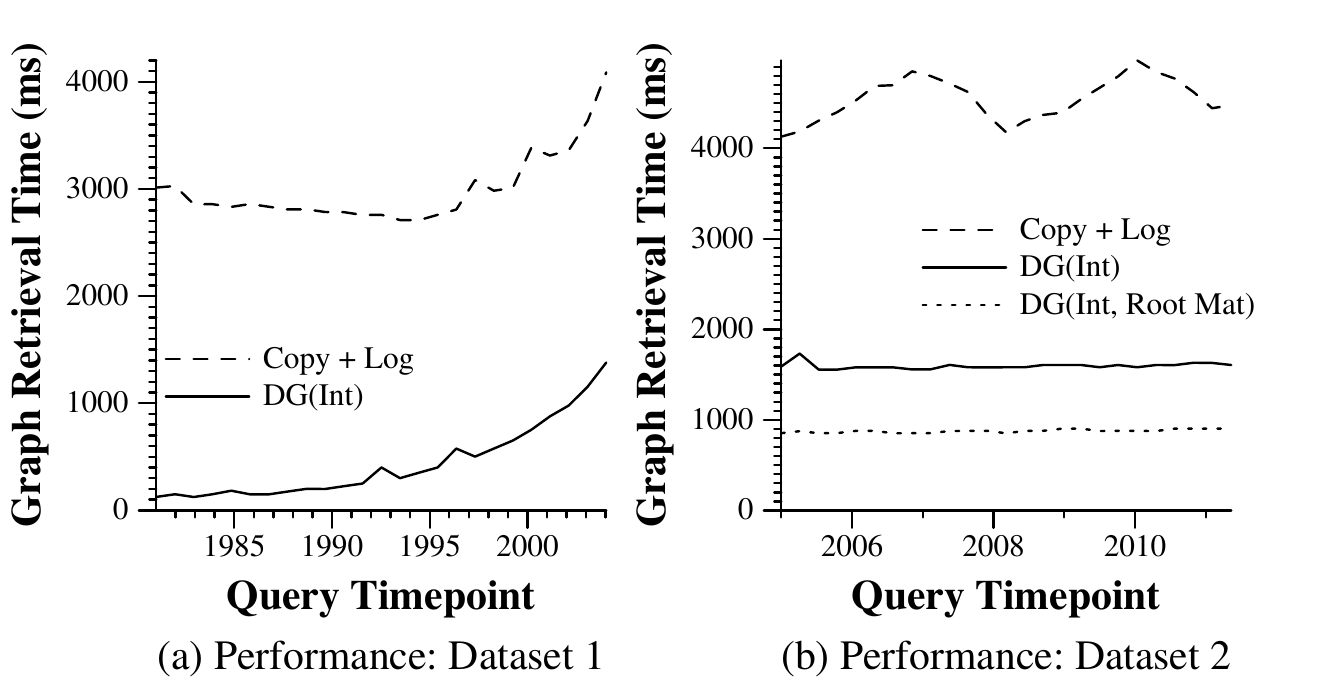}
\caption{Comparing DeltaGraph and Copy+Log. Int and Bal denote the Intersection and Balanced functions respectively.}
\label{fig:CLvDG}
\end{figure}

Figure~\ref{fig:CLvDG} shows the results of our comparison between Copy+Log and DeltaGraph
approaches for time taken to execute 25 uniformly spaced queries on Datasets 1 and 2. The leaf-eventlist
sizes were chosen so that the disk storage space consumed by both the approaches was about the same.
For similar disk space constraints (450MB and 950MB for Dataset 1 and 2, respectively), the DeltaGraph could afford a smaller size of $L$ and hence higher number of snapshots than
the Copy+Log approach. As we can see, the best DeltaGraph variation outperformed the Copy+Log
approach by a factor of at least 4, and orders of magnitude in several cases.



\begin{figure}[t]
\centering
 \includegraphics [width=0.5\textwidth] {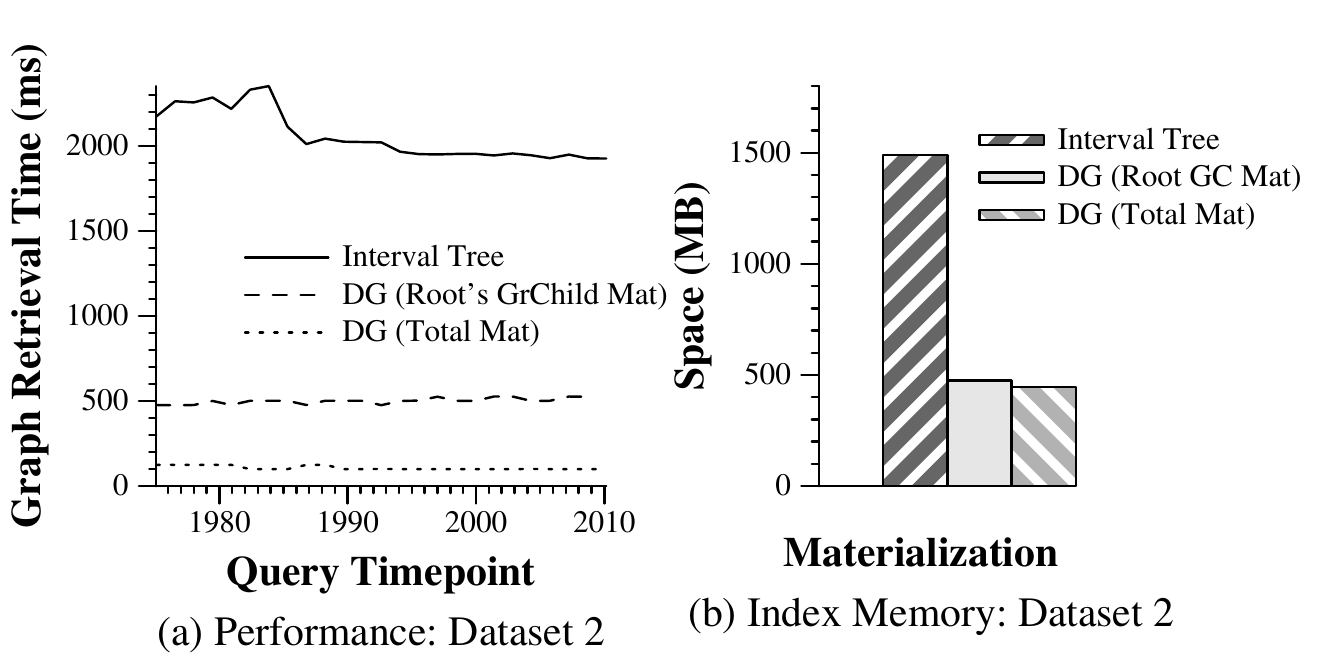}
\caption{Performance of different DeltaGraph configurations vs. Interval Tree}
\label{fig:ITvDG}
\end{figure}

Figure~\ref{fig:ITvDG} shows the comparison between an in-memory interval tree and two DeltaGraph
variations: (1) with low materialization,  (2) with all leaf nodes materialized. We compared these configurations
for time taken to execute 25 queries on Dataset 2, using $k=4$ and $L=30000$. We can see 
that both the DeltaGraph approaches outperform the interval
tree approach, while using significantly less memory than the interval tree (even with total materialization). The largely disk-resident DeltaGraph with root's grandchildren materialized is more than four times as fast as the regular
approach, whereas the total materialization approach, a more fair comparison, is much faster.

We also evaluated a naive approach similar to the Log technique, with raw events being read from input files directly (not shown in graphs). The average retrieval times were worse than DeltaGraph by factors of 20 and 23 for Datasets 1 and 2 respectively.

\begin{figure}[t]
\centering
  \includegraphics [width=0.5\textwidth] {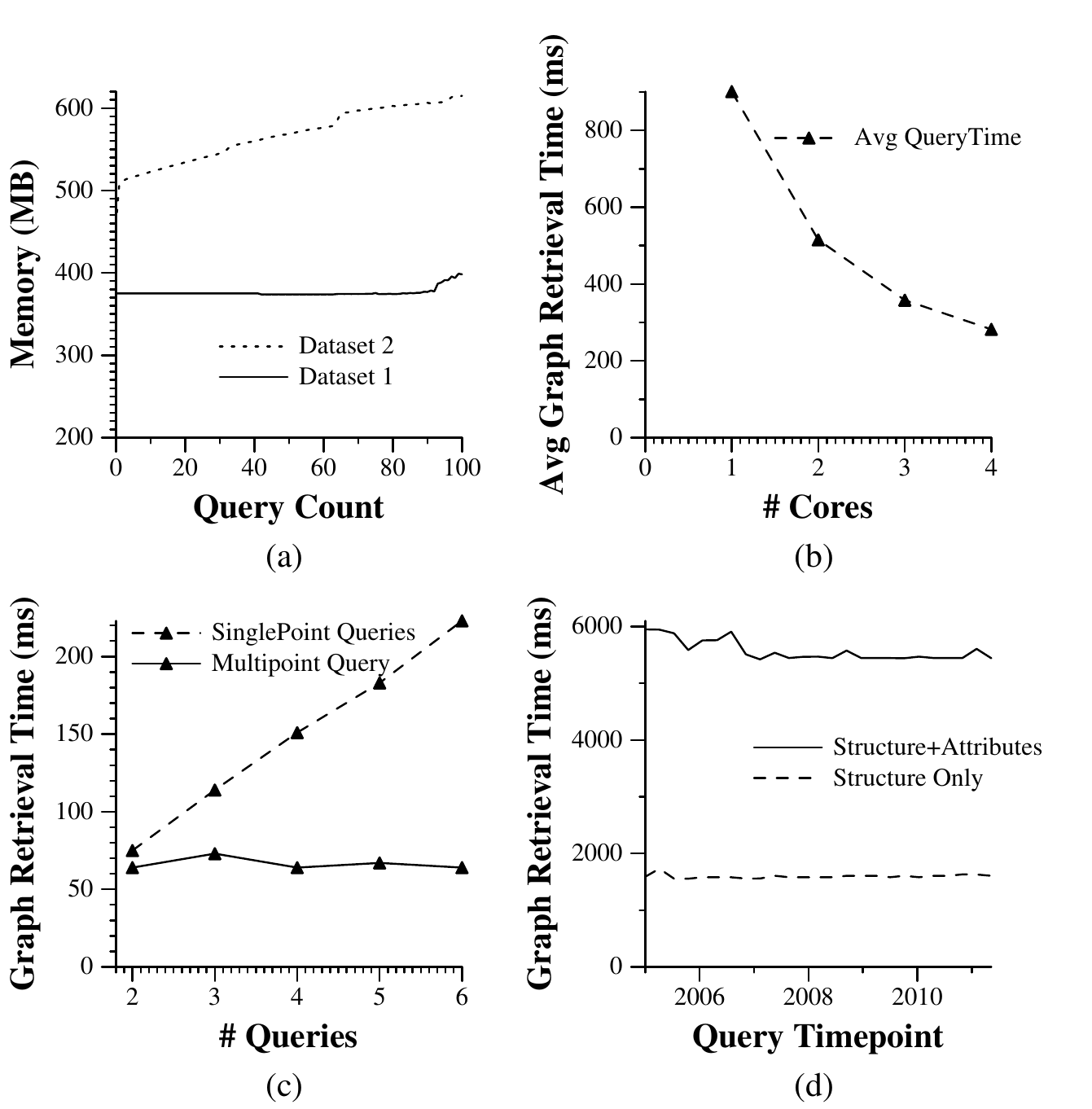}
  \caption{(a) Cumulative GraphPool memory consumption; (b) Multi-core parallelism (Dataset 2); (c) Multipoint query execution vs multiple singlepoint queries; (d) Retrieval with and without attributes (Dataset 2)}
\label{fig:GPMem}
\end{figure}

\topic{GraphPool memory consumption:}
Figure~\ref{fig:GPMem}(a) shows the total (cumulative) memory consumption of
the GraphPool when a sequence of 100 singlepoint snapshot retrieval queries, uniformly spaced over the
life span of the network, is executed against Datasets 1 and 2.
By exploiting the overlap between these snapshots, the GraphPool is able to maintain a large number
of snapshots in memory. 
For Dataset 2, if the 100 graphs 
were to be stored disjointly, the total memory consumed would be
50GB, whereas the GraphPool only requires about 600MB. 
The plot of Dataset 1 is almost a constant because, for this dataset, any historical snapshot is a
subset of the current graph.
The minor increase toward the end is due to the increase in the bitmap size, required to 
accommodate new queries. 

\topic{Multicore Parallelism:} Figure~\ref{fig:GPMem}(b) shows the advantage of concurrent query processing on a multi-core processor using a partitioned DeltaGraph approach, 
where we retrieve the graph parallely using multiple threads. We observe near-linear speedups further validating our parallel design.

\topic{Multipoint queries:}
Figure~\ref{fig:GPMem}(c) shows the time taken to retrieve multiple graphs using our multipoint
query retrieval algorithm, and multiple invocations of the single query retrieval algorithm on
Dataset 1. The x-axis represents the number of snapshot queries, which were chosen to be 1 month
apart. As we can see, the advantages of multipoint query retrieval are quite significant because of
the high overlap between the retrieved snapshots.
%

\topic{Advantages of columnar storage:} Figure \ref{fig:GPMem}(d) shows the performance benefits of
our columnar storage approach for Dataset~1. As we can see, if we are only interested in the network structure,
our approach can improve query latencies by more than a factor of 3. 

\topic{Bitmap penalty:}
We compared the penalty of using the bitmap filtering procedure in GraphPool, by doing a PageRank
computation without and with use of bitmaps. We observed that the execution time increases from
1890ms to 2014ms, increase of less than 7\%.


\label{subsec:sysperf}
\begin{figure}[t]
\centering
  \includegraphics [width=0.5\textwidth] {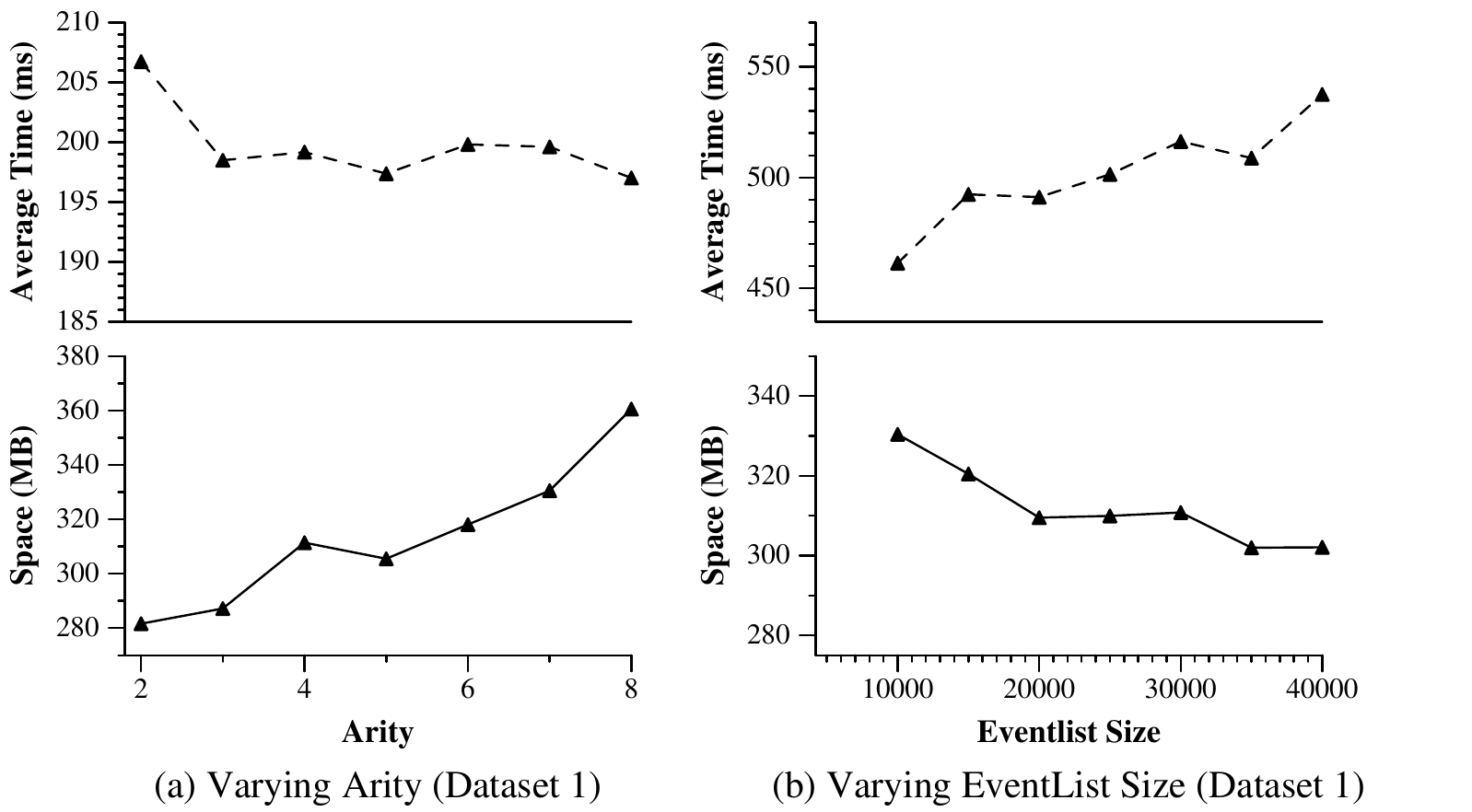}
\caption{Effect of varying arity and leaf-eventlist sizes}
\label{fig:ELandAR}
\end{figure}

\topic{Effect of DeltaGraph construction parameters:}
We measured the average query times and storage space consumption for different values of the 
arity ($k$) and leaf-eventlist sizes ($L$) for Dataset 1. Figure~\ref{fig:ELandAR}(a) shows that with an increase
in the arity of the DeltaGraph, the average query time decreases rapidly in the beginning, but
quickly flattens. 
On the other hand, the space requirement increases in general with small exceptions when
the height of the tree does not change with increasing arity. We omit a detailed
discussion on the exceptions as their impact is minimal. Again, referring back to the discussion in
Section~\ref{subsec:difffunct}, the results corroborate our claim that higher arity leads to
smaller DeltaGraph heights and hence smaller query times, but results in higher space requirements. 
The effect of the leaf-eventlist size is also as expected. As the leaf-eventlist size increases,
the total space consumption decreases (since there are fewer leaves), but the query times also
increase dramatically.

\begin{figure}[t]
  \includegraphics [width=0.5\textwidth] {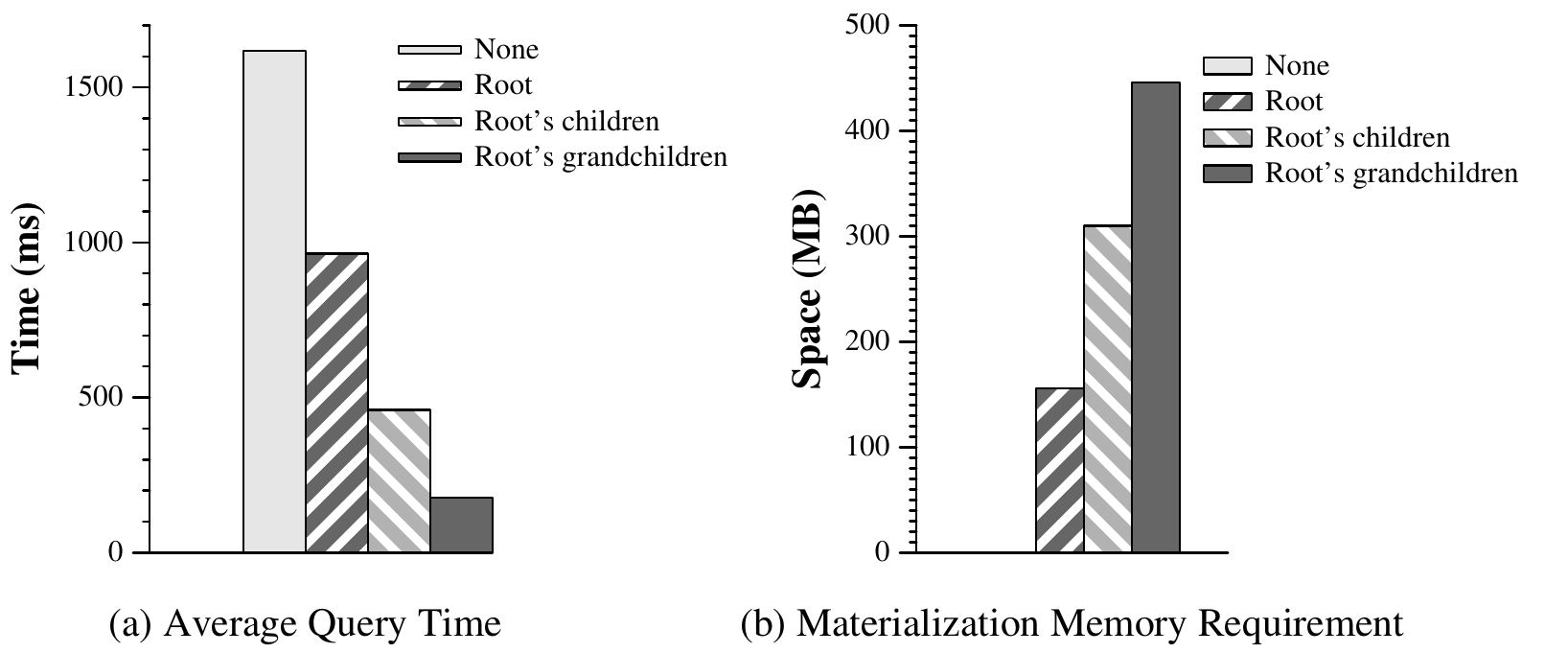}
\caption{Effect of materialization} 
\label{fig:Mat}
\end{figure}

\topic{Materialization:}
Figure~\ref{fig:Mat} shows the benefits of materialization for a DeltaGraph and the
associated cost in terms of memory, for Dataset 2 with arity = 4 and using the Intersection differential
function. 
We compared four different situations: (a) no materialization, (b) root materialized, (c) both
children of the root materialized, and (d) all four grandchildren of the root materialized. 
The results are as expected -- we can significantly reduce the query latencies (up to a factor of 8) at
the expense of higher memory consumption.

\begin{figure}[t]
\centering
  \includegraphics [width=0.5\textwidth] {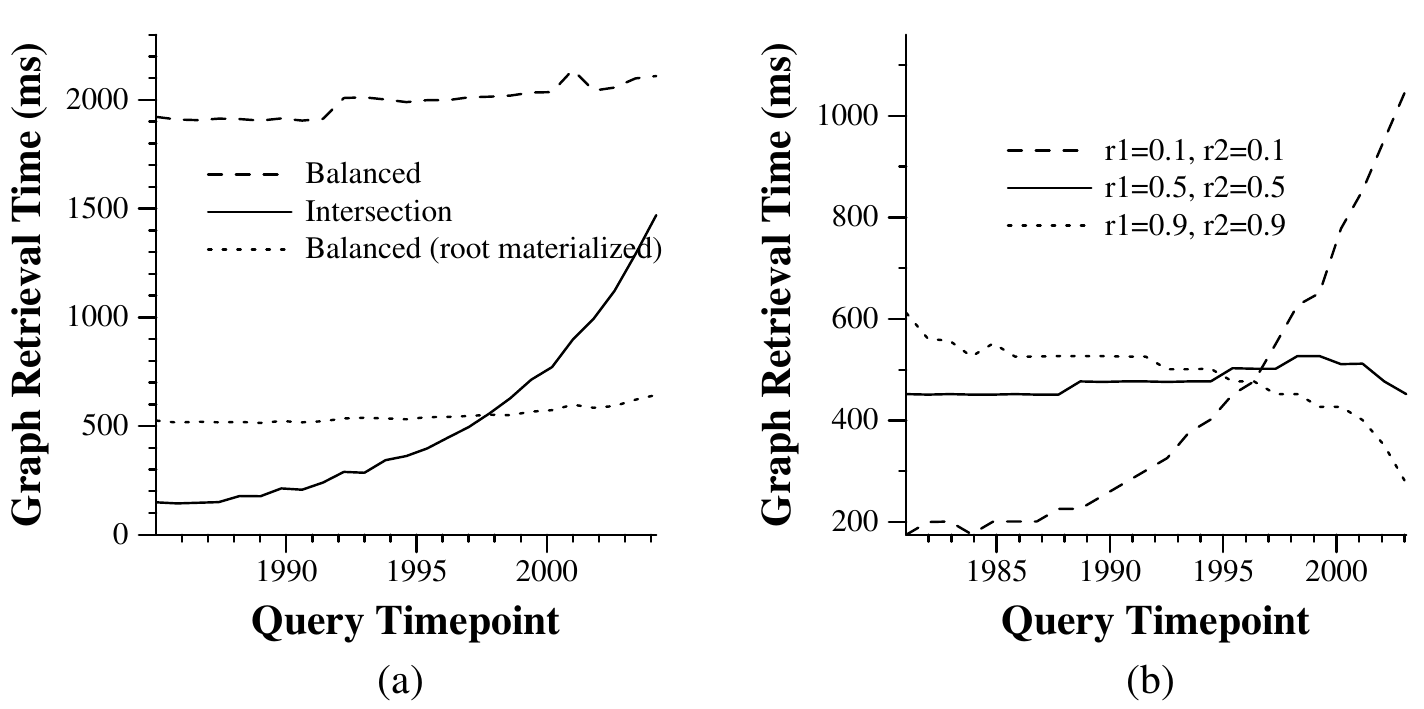}
\caption{(a) Comparison of differential functions, Intersection vs Balanced (Dataset 1) and (b) Comparison of different Mixed function configurations (Dataset 1) }
\label{fig:FUN}
\end{figure}

\topic{Differential functions:}
In Section~\ref{subsec:difffunct}, we discussed how the choice of an appropriate differential
function can help achieve desired distributions of retrieval times for a given network.
Figure~\ref{fig:FUN}(a) compares the behavior of Intersection and Balanced functions with and without
root materialization on Dataset 1. On the growing-only graph, using the Intersection function
results in skewed query times, with the larger (newer) snapshots taking longer to load. 
The balanced function, on the other hand, provides a
more uniform access pattern, although the average time taken is higher. By materializing the root
node, the average becomes comparable to that of Intersection, yielding a uniform access pattern.
A few different configurations for the mixed function are shown in Figure~\ref{fig:FUN}(b), with 
$r_{1}=0.5$, $r_{2}=0.5$ denoting the Balanced function. As we can see, by choosing an appropriate
differential function, we can exercise fine-grained control over the query retrieval times thus
validating one of our main goals with the DeltaGraph approach.

\section{Conclusions and Future Work}
\label{sec:conclusion}
In this paper, we presented an approach for managing historical data for large information networks,
and for executing snapshot retrieval queries on them. We presented
DeltaGraph, a distributed hierarchical structure to compactly store the historical trace of a network, and
GraphPool, a compact way to maintain and operate upon multiple graphs in memory. Our experimental evaluation shows that
the choice of DeltaGraph is superior to the existing alternatives. We showed both analytically
and empirically that the flexibility of DeltaGraph helps control the distribution of query access
times through appropriate parameter choices at construction time, and memory materialization at
runtime. Our experimental evaluation demonstrated the impact of many of our optimizations, including
multi-query optimization and columnar storage. Our work so far has also opened up many
opportunities for further work, including developing improved DeltaGraph construction algorithms and 
techniques for processing different types of temporal queries over the historical trace, that we are
planning to pursue in future work.

{
\bibliographystyle{plain}
\bibliography{historical}
}
\end{document}